\newread\epsffilein    
\newif\ifepsffileok    
\newif\ifepsfbbfound   
\newif\ifepsfverbose   
\newdimen\epsfxsize    
\newdimen\epsfysize    
\newdimen\epsftsize    
\newdimen\epsfrsize    
\newdimen\epsftmp      
\newdimen\pspoints     
\def\epsfbox#1{\global\def\epsfllx{72}\global\def\epsflly{72}%
   \global\def\epsfurx{540}\global\def\epsfury{720}%
   \def\lbracket{[}\def\testit{#1}\ifx\testit\lbracket
   \let\next=\epsfgetlitbb\else\let\next=\epsfnormal\fi\next{#1}}%
\def\epsfgetlitbb#1#2 #3 #4 #5]#6{\epsfgrab #2 #3 #4 #5 .\\%
   \epsfsetgraph{#6}}%
\def\epsfnormal#1{\epsfgetbb{#1}\epsfsetgraph{#1}}%
\def\epsfgetbb#1{%
\openin\epsffilein=#1
\ifeof\epsffilein\errmessage{I couldn't open #1, will ignore it}\else
   {\epsffileoktrue \chardef\other=12
    \def\do##1{\catcode`##1=\other}\dospecials \catcode`\ =10
    \loop
       \read\epsffilein to \epsffileline
       \ifeof\epsffilein\epsffileokfalse\else
          \expandafter\epsfaux\epsffileline:. \\%
       \fi
   \ifepsffileok\repeat
   \ifepsfbbfound\else
    \ifepsfverbose\message{No bounding box comment in #1; using defaults}\fi\fi
   }\closein\epsffilein\fi}%
\def\epsfsetgraph#1{%
   \epsfrsize=\epsfury\pspoints
   \advance\epsfrsize by-\epsflly\pspoints
   \epsftsize=\epsfurx\pspoints
   \advance\epsftsize by-\epsfllx\pspoints
   \epsfxsize\epsfsize\epsftsize\epsfrsize
   \ifnum\epsfxsize=0 \ifnum\epsfysize=0
      \epsfxsize=\epsftsize \epsfysize=\epsfrsize
     \else\epsftmp=\epsftsize \divide\epsftmp\epsfrsize
       \epsfxsize=\epsfysize \multiply\epsfxsize\epsftmp
       \multiply\epsftmp\epsfrsize \advance\epsftsize-\epsftmp
       \epsftmp=\epsfysize
       \loop \advance\epsftsize\epsftsize \divide\epsftmp 2
       \ifnum\epsftmp>0
          \ifnum\epsftsize<\epsfrsize\else
             \advance\epsftsize-\epsfrsize \advance\epsfxsize\epsftmp \fi
       \repeat
     \fi
   \else\epsftmp=\epsfrsize \divide\epsftmp\epsftsize
     \epsfysize=\epsfxsize \multiply\epsfysize\epsftmp   
     \multiply\epsftmp\epsftsize \advance\epsfrsize-\epsftmp
     \epsftmp=\epsfxsize
     \loop \advance\epsfrsize\epsfrsize \divide\epsftmp 2
     \ifnum\epsftmp>0
        \ifnum\epsfrsize<\epsftsize\else
           \advance\epsfrsize-\epsftsize \advance\epsfysize\epsftmp \fi
     \repeat     
   \fi
   \ifepsfverbose\message{#1: width=\the\epsfxsize, height=\the\epsfysize}\fi
   \epsftmp=10\epsfxsize \divide\epsftmp\pspoints
   \vbox to\epsfysize{\vfil\hbox to\epsfxsize{%
      \includegraphics{#1}%
      \hfil}}%
\epsfxsize=0pt\epsfysize=0pt}%

{\catcode`\%=12 \global\let\epsfpercent=
\long\def\epsfaux#1#2:#3\\{\ifx#1\epsfpercent
   \def\testit{#2}\ifx\testit\epsfbblit
      \epsfgrab #3 . . . \\%
      \epsffileokfalse
      \global\epsfbbfoundtrue
   \fi\else\ifx#1\par\else\epsffileokfalse\fi\fi}%
\def\epsfgrab #1 #2 #3 #4 #5\\{%
   \global\def\epsfllx{#1}\ifx\epsfllx\empty
      \epsfgrab #2 #3 #4 #5 .\\\else
   \global\def\epsflly{#2}%
   \global\def\epsfurx{#3}\global\def\epsfury{#4}\fi}%
\def\epsfsize#1#2{\epsfxsize}
\let\epsffile=\epsfbox

\documentstyle [12pt,twoside]{article}
\oddsidemargin=0in
\evensidemargin=0in
\topmargin=0in
\headheight=0in
\headsep=0in
\textheight=9in
\footheight=3ex
\footskip=4ex
\textwidth=6in
\hsize=6in
\parskip=0pt
\lineskip=0pt
\begin{document}
\small
\par\noindent{\large\bf Phase mixing in time-independent Hamiltonian systems}
\vskip .15in
\par\noindent{\large Henry E. Kandrup$^{*}$}
\vskip .20in
\par\noindent{\it Departments of Astronomy and Physics and Institute for 
Fundamental Theory, University of Florida, Gainesville, Florida 32611}
\vskip .25in
\centerline{(Accepted 1998 \hfill . Received 1998 \hfill; in original
form 1998 \hfill) }
\vskip .2in
\par\noindent
{\bf ABSTRACT}
\par\noindent
This paper describes the evolution of localised ensembles of initial
conditions in two- and three-dimensional time-independent potentials
which admit a coexistence of regular and chaotic orbits. The coarse-grained 
approach towards an invariant, or near-invariant, distribution was probed by 
tracking (1) moments 
${\langle}x^{i}y^{j}z^{k}p_{x}^{l}p_{y}^{m}p_{z}^{n}{\rangle}$ for 
$i+j+k+l+m+n{\;}{\le}{\;}4$ and (2) binned representations of reduced 
distributions $f(Z_{a},Z_{b})$ for $a{\;}{\ne}{\;}b=x,y,z,p_{x},p_{y},p_{z}$
computed at fixed intervals ${\Delta}t$. For ensembles of ``unconfined'' 
chaotic orbits in two-dimensional systems not stuck near islands by
cantori, the moments evolve exponentially. Quantities like the dispersion 
${\sigma}_{px}$, which start small and eventually asymptote towards a larger 
value, initially grow exponentially in time at a rate comparable to 
${\overline \chi}$, the mean value of the largest short time Lyapunov 
exponent for orbits in the ensemble. Quantities like 
$|{\langle}p_{x}{\rangle}|$, that can start large but eventually asymptote 
towards zero, decrease exponentially. With respect to a discrete $L^{p}$ norm, 
reduced distributions $f(Z_{a},Z_{b})$ 
generated from successive snapshots exhibit an overall exponential decay 
towards a near-invariant $f_{niv}(Z_{a},Z_{b})$, although a plot of 
$Df(t){\;}{\equiv}{\;}||f(t)-f_{niv}||$ can exhibit considerable structure. 
Implementing an additional coarse-graining by averaging over several 
successive snapshots can reduce the amount of structure and increase the rate 
${\Lambda}_{ab}$ at which ensembles asymptote towards $f_{niv}$. Regular 
ensembles behave very differently, both moments and $Df$ evolving in a fashion 
better represented by a power law time dependence. ``Confined'' chaotic orbits,
initially stuck near regular islands because of cantori, exhibit an 
intermediate behaviour. The behaviour of ensembles evolved in 
three-dimensional potentials is qualitatively similar, except that, in this 
case, it is relatively likely to find one direction in configuration space 
which is ``less chaotic'' than the other two, so that quantities like
${\Lambda}_{ab}$ depend more sensitively on which phase space variables 
one tracks.
\vfill
\par\noindent
$^{*}$E-mail: kandrup@astro.ufl.edu
\vfill\eject
\vskip .2in
\par\noindent{\bf 1 INTRODUCTION AND MOTIVATION}
\vskip .1in
\par\noindent
Galactic dynamicists are interested in how, given generic
initial conditions, a self-gravitating system of nearly point mass stars
will evolve towards a statistical equilibrium or near-equilibrium. This 
question can be, and in many cases has been, formulated in the context of 
the full many-body problem, where the natural arena of physics is the 
$6N$-dimensional phase space. However, at least for large $N$ it is 
customary to assume that the system can be characterised by a one-particle
distribution function, $F(t)$, satisfying the collisionless Boltzmann 
equation, which is believed to describe the $N$-body problem correctly in a 
suitable $N\to\infty$ limit (see, e.g., Binney \& Tremaine 1987). 

In this latter setting, conventional wisdom is dominated by Lynden-Bell's 
(1967) theory of violent relaxation, which views the evolution of the 
one-particle $F$ as a phase mixing process. A mathematically rigorous analysis 
of this phase mixing must involve a description which incorporates the fact 
that the collisionless Boltzmann equation is an infinite-dimensional 
Hamiltonian system, with $F$, the fundamental dynamical variable, evolving in 
the infinite-dimensional phase space of distribution functions (see, e.g., 
Morrison 1980, Kandrup 1998, and references contained therein). However, 
dating back at least to Lynden-Bell's original paper (cf.~his balls rolling in 
a pig-trough), there has been the expectation that the qualitative character 
of the approach towards a statistical equilibrium, i.e., a time-independent 
solution to the collisionless Boltzmann equation, can be understood 
heuristically in the context of a much simpler toy problem, namely a 
one-particle distribution function evolved in a fixed time-independent 
potential. 

This seems quite reasonable. However, the question then arises as to whether 
that potential is integral or near-integrable, so that (almost) all the orbits
are regular, or whether instead it is far from integrable, so that the phase
space admits a coexistence of large measures of both regular and chaotic 
orbits. If the orbits are all regular, as was implicit in Lynden-Bell's 
original discussion, any tendency for $F$ to evolve towards an equilibrium 
should be relatively weak. In particular, one might expect that, as probed by 
the behaviour of coarse-grained distribution functions or various moments 
associated with $F$, any approach towards equilibrium will proceed as a power 
law in time. By contrast, if the initial $F(t_{0})$ samples a phase space 
region where all the orbits are chaotic, one might expect a much more 
efficient evolution towards a (near-)equilibrium, the approach now proceeding 
exponentially in time.

The objective of the research described in this paper was to understand the 
bulk properties of flows in a complex time-independent potential which admits 
a coexistence of significant measures of both regular and chaotic orbits. Is
it, for example, true that ensembles of chaotic orbits exhibit exponential 
phase mixing, whereas regular ensembles exhibit a more modest power law phase 
mixing? 
This is a problem of obvious interest in its own right. However, 
this work should also be understood as a prolegomena to a more complete 
treatment of violent relaxation which will generalise this work in two 
important ways, namely by allowing for self-consistency (Habib, Kandrup, 
Pogorelov, \& Ryne 1998) and discreteness effects (Pogorelov \& Kandrup 1998). 
To assess the importance of self-consistency and discreteness effects, 
one first needs to understand what happens when these complications are
completely ignored. 

Self-consistency must eventually be
incorporated because the collisionless Boltzmann equation describes an 
evolution in response to a self-consistently determined potential. Although
less obvious, discreteness effects could also play an important role.
Indeed, numerical computations have shown that even very weak friction and 
noise, which are often used to model discreteness effects (see, e.g., 
Chandrasekhar 1943), can have significant evolutionary effects on a 
surprisingly short time scale (Habib, Kandrup, \& Mahon 1997, 1998, Kandrup 
1999) by facilitating diffusion through cantori (Aubry \& Andre 1978, Mather 
1982) or along an Arnold web (Arnold 1964).

The numerical experiments described here involved specifying an initial 
distribution, $F({\bf r},{\bf p},t_{0})$, localised in a small phase space 
region, and understanding how $F$ changes when evolved into the future. In 
particular, is there an efficient approach towards an invariant, or 
near-invariant, distribution? As a practical matter, this was done   
by (1) sampling $F(t_{0})$ to obtain a set of initial conditions, 
(2) integrating these initial conditions into the future, and then (3) 
analysing the orbits to extract systematic trends.

In this analysis two specific diagnostics played an important role:
\par\noindent
1. {\it Coarse-grained reduced distribution functions}. Obtaining a high 
resolution
numerical representation of the full four- or six-dimensional distribution
function is very expensive computationally. Even in four dimensions, generating
enough orbits to justify a matrix with as many as $40$ cells on a side is
very time-consuming, except on the very fastest computers. Indeed, such a 
matrix would contain $2.56\times 10^{6}$ cells, so that even storing the data 
for a large number of snapshots would be a nontrivial concern. For this 
reason, attention here focused on reduced two-variable distributions 
$f(Z_{a},Z_{b})$, with $Z_{a}{\;}{\ne}{\;}Z_{b}=x,y,z,p_{x},p_{y},p_{z}$, the 
obvious question being whether such reduced $f$'s evolve towards a 
time-independent form and, if so, how this evolution proceeds in time.
\par\noindent
2. {\it Moments of the four- or six-dimensional distribution function}. 
Following
a set of moments, even working up to relatively high order, entails tracing a
relatively small collection of numbers, rather than phase space functions,
and is thus comparatively inexpensive. Moreover, moments are often easily
related to observables, so that a knowledge of their form allows one to 
determine whether, in terms of real, measureable quantities, orbit ensembles 
evolve towards a time-independent state. 

As stressed already, one crucial issue in all this is: how does the 
qualitative character of the flow depend on whether the initial $F(t_{0})$ 
probes phase space regions characterised by regular or chaotic orbits? Even 
considering only chaotic initial conditions, one might expect to see very
different sorts of behaviour. For example, in two-dimensional Hamiltonian 
systems one anticipates important distinctions between confined, or sticky, 
regions (cf. Contopoulos 1971), corresponding to orbits initially trapped by
cantori near regular islands, and unconfined, or nonsticky, regions with 
orbits that move relatively unimpeded through large portions of the chaotic 
regions far from any regular island. However, the average divergence of nearby 
trajectories is set by various short time Lyapunov exponents, the typical 
values of which will differ for sticky and non-sticky regions (Mahon, 
Abernathy, Bradley, \& Kandrup 1995), so that one 
might expect correlations between the values of short time Lyapunov exponents 
and the bulk properties of initially localised orbit ensembles.

Earlier work by Kandrup \& Mahon (1994) and Mahon, Abernathy, Bradley, \& 
Kandrup (1995) provided some tentative conclusions about two-dimensional
Hamiltonian systems. For example, it was found that, with respect to a 
discrete $L^{1}$ form, binned distributions $f(x,y)$ and $f(p_{x},p_{y})$ 
associated with chaotic ensembles of fixed energy typically exhibit a 
coarse-grained, exponential approach towards a near-invariant distribution 
$f_{niv}$ at a rate ${\Lambda}$ that correlates with ${\overline\chi}$, the 
mean positive short time Lyapunov exponent for the ensemble. This distribution 
is near-invariant in the sense that, once achieved, it only changes on 
significantly longer times. However, $f_{niv}$ does not in general correspond 
to a {\it true} invariant distribution $f_{iv}$. When integrated for much 
longer times, $f$ changes as orbits slowly diffuse through cantori to 
access regions that were avoided systematically over shorter time scales. The 
{\it true} invariant distribution, which $f$ may approach only at very late 
times, corresponds to a microcanonical equilibrium, i.e., a uniform (in 
canonical coordinates) population of those phase space regions on the 
constant energy surface that are accessible to the ensemble.  

That earlier work involved three distinct coarse-grainings, namely: (i) 
considering reduced distributions $f(Z_{a},Z_{b})$, rather than the full $F$; 
(ii) considering binned representations of $f$; and (iii) averaging over 
several successive snapshots (this to reduce the number of orbits to be
computed!). Coarse graining {\it per se} is not a bad idea since, given 
Liouville's Theorem, there can be no pointwise approach towards a 
time-independent equilibrium. However, one may wonder whether both temporal 
and phase space coarse grainings are necessary. In particular, it would seem 
important to determine whether the conclusions of Kandrup \& Mahon (1994) 
remain valid in the absence of any temporal coarse graining. The first two 
coarse-grainings are reasonable physically in analysing experimental data 
where, at a fixed moment in time, one makes a measurement of some specific 
observable with finite phase space resolution. Temporal coarse-grainings would 
seem less well motivated physically. Another obvious question, as yet 
unanswered, is whether, for a given ensemble, the convergence rate 
${\Lambda}_{ab}$ is the same for different reduced distributions 
$f(Z_{a},Z_{b})$. Different choice of variables could lead to different values
of ${\Lambda}_{ab}$ since, in general, the initial ensemble could ``mix'' at 
different rates in different directions. 

Merritt \& Valluri (1996) subsequently exploited techniques similar to those
developed by Kandrup \& Mahon (1994) and Mahon, Abernathy, Bradley, \& Kandrup
(1995) to conclude that, in at least some three-dimensional potentials, 
ensembles of chaotic initial conditions will again exhibit an exponential 
approach towards a near-invariant $f_{niv}$. Unlike earlier work on 
two-dimensional systems, that paper did not implement a temporal 
coarse-graining.
However, the authors probed the approach towards $f_{niv}$ using a diagnostic 
which yields no information about direction-dependence, so that they could not 
address the possibility of ``mixing'' at different rates in different 
directions. Moreover, none of the earlier work on either two- or 
three-dimensional flows has addressed the question of how the lower order 
moments evolve in time.

To a large extent, these moments divide into two types, namely (1) those like 
${\sigma}_{px}$, the dispersion in $p_{x}$, which start small and presumably 
asymptote towards a larger nonzero value, and (2) those like the mean 
${\langle}p_{x}{\rangle}$ which typically start with a finite value but, at 
least for chaotic ensembles, would be expected to evolve towards zero. It 
seems reasonable to conjecture that, for moments of the first type, 
chaotic ensembles exhibit an initial growth that is exponential in time, 
whereas regular ensembles exhibit a slower, power law growth. Alternatively, 
one might anticipate that moments like ${\langle}p_{x}{\rangle}$ will decay to 
zero exponentially for chaotic ensembles but -- if at all -- only much more 
slowly for regular ensembles. 

Section 2 describes the behaviour of moments and reduced distributions 
extracted from an analysis of orbits in two-dimensional potentials. Section
3 then discusses how these results are changed by allowing for a 
three-dimensional potential. Section 4 concludes by summarising the principal
results and commenting on their significance.
\vskip .15in
\par\noindent{\bf 2 TWO-DIMENSIONAL SYSTEMS}
\vskip .1in
\par\noindent{\bf 2.1 Description of what was computed}
\vskip .1in
\par\noindent
The conclusions described in this Section derive from a 
study of three different Hamiltonian systems of the form
$$H={1\over 2}{\Bigl(}p_{x}^{2}+p_{y}^{2}{\Bigr)}+V(x,y). \eqno(1) $$
The potentials that were used included the sixth order truncation of the 
Toda (1967) potential, namely
$$V(x,y)={1\over 2}{\Bigl(}x^{2}+y^{2}{\Bigr)}+x^{2}y-{1\over 3}y^{3}
+{1\over 2}x^{4}+x^{2}y^{2}+{1\over 2}y^{4}$$
$$+x^{4}y+{2\over 3}x^{2}y^{3}-{1\over 3}y^{5}
+{1\over 5}x^{6}+x^{4}y^{2}+{1\over 3}x^{2}y^{4}+{11\over 45}y^{6};
\eqno(2) $$
the sum of isotropic and anisotropic softened Plummer potentials, namely
(Mahon, Abernathy, Bradley, \& Kandrup 1995)
$$V(x,y)=-{1\over {\Bigl(}c^{2}+x^{2}+y^{2}{\Bigr)}^{1/2}}
-{m\over {\Bigl(}c^{2}+x^{2}+ay^{2}{\Bigr)}^{1/2}}, \eqno(3) $$
\par\noindent
with $c=20^{2/3}{\;}{\approx}{\;}0.136$, $a=0.1$ and $m=0.3$;
and the so-called dihedral potential of Armbruster, Guckenheimer, \&
Kim (1989) for one particular set of parameter values, namely
$$V(x,y)=-(x^{2}+y^{2})+{1\over 4}
(x^{2}+y^{2})^{2}-{1\over 4}x^{2}y^{2} . \eqno(4) $$
Although these potentials manifest very different symmetries, the basic 
conclusions about the evolution of orbit ensembles are very similar, which 
suggests that they are robust. Because of this similarity, the 
discussion here focuses primarily on the dihedral potential which, being the 
most inexpensive to implement computationally, could be explored for the 
largest number of orbits, thus yielding results of the greatest statistical 
significance.

Each experiment involved an ensemble of $81\times 81=6561$ initial conditions
with the same fixed energy $E$. In general, different ensembles were chosen to 
sample only regular or only chaotic orbits, as identified using a surface of 
section. These were generated by uniformly sampling a square in the $y-p_{y}$ 
plane, typically with sides ${\Delta}y={\Delta}p_{y}=0.2$, setting $x=0$, and 
solving for $p_{x}=p_{x}(y,p_{y},E)>0$. A Burlisch-Stoer or fourth order 
Runge-Kutta intergrator was used to evolve each initial condition for 
$t{\;}{\ge}{\;}256$, a time corresponding to an interval of order 
$100-200$ crossing times $t_{cr}$. The integrator solved simultaneously
for the evolution of a small, linearised perturbation, renormalised at fixed
intervals ${\Delta}t=1.0$, so as to obtain an estimate of the largest 
Lyapunov exponent ${\chi}$ (see, e.g., Lichtenberg \& Lieberman 1992). 
The phase space coordinates and short time ${\chi}$ were recorded for each 
orbit at intervals ${\Delta}t=0.25$, $0.5$, or $1.0$. Most of the analysis 
focused on ensembles of chaotic orbits, which were expected to exhibit the 
most interesting behaviour.

Each ensemble of initial conditions was interpreted as sampling an initial 
distribution $F(x,y,p_{x},p_{y},t_{0})$, and the evolved orbits were 
interpreted as yielding an approximation to the true $F(t)$ associated with 
the initial $F(t_{0})$. The data were thus binned to yield $n \times n$ 
coarse-grained gridded approximations of the six possible reduced 
two-dimensional distributions $f(Z_{a},Z_{b},t)$ with 
$a{\;}{\ne}{\;}b=x,y,p_{x},p_{y}$. Most of the analysis involved 
the choice $n=20$. It was verified that $n=10$, $30$, and $40$ yielded no 
differences that could not be attributed to finite number statistics.

The expected coarse-grained approach towards equilibrium associated with 
chaotic ensembles was probed in two distinct ways. The first involved 
determining how, with respect to an appropriate norm, the coarse-grained 
distributions $f(t)$ evolve towards an invariant, or near-invariant, 
$f_{inv}$. Consistent with earlier experiments (Kandrup \& Mahon, 1994, Mahon, 
Abernathy, Bradley, \& Kandrup 1995), it was observed that, after a relatively 
short interval, the orbit ensembles typically evolved towards a state which is 
near-invariant in the sense that, once attained, it only exhibits large 
systematic shifts on a much longer time scale. Physically, this state seems to 
correspond to a near-constant population (in canonical coordinates) of those 
portions of the constant energy hypersurface that are easily accessible to the 
orbits, i.e., not blocked by relatively impenetrable cantori. The evolution 
observed at much later times involves orbits diffusing through cantori to 
probe regions on the constant energy surface that were avoided systematically 
on short time scales (cf. MacKay, Meiss, \& Percival 1984). Approximations to 
the near-invariant distributions $f_{niv}(Z_{a},Z_{b})$ were generated by 
averaging over orbital data generated from $64$ successive snapshots at 
intervals ${\Delta}t=1.0$, usually extending from $t=193$ to $t=256$ but, when 
convergence was slow, from a significantly later interval. The approach of 
$f(t)$ towards $f_{niv}$ was quantified by tracking the ``distance'' between 
$f(t)$ and $f_{niv}$, as defined by the discrete $L^{p}$ norm
$$Df(Z_{a},Z_{b},t)=
{\Biggl(} {\sum_{a}\sum_{b}|f(Z_{a},Z_{b},t)-f_{niv}(Z_{a},Z_{b})|^{p} 
\over \sum_{a}\sum_{b}|f_{niv}(Z_{a},Z_{b})|^{p} }{\Biggr)}^{1/p}, \eqno(5) $$
with $p=1$ or $2$. 

The orbital data were also used to track the evolution of all possible moments
${\langle}x^{i}y^{j}p_{x}^{k}p_{y}^{l}{\rangle}$ with non-negative integers
$i+j+k+l{\;}{\le}{\;}4$. One can view these moments either directly as probes 
of the $6561$-orbit ensemble or as probes of statistics of the smooth phase 
space distribution $F(x,y,p_{x},p_{y},t)$. The latter viewpoint suggests that 
one might expect correlations between the way in which $f(t)$ evolves towards 
$f_{niv}$ and the way in which such moments as ${\langle}p_{x}{\rangle}$ 
evolve towards zero. In particular, if $Df(t)\to 0$ exponentially, one might 
expect that ${\langle}p_{x}(t){\rangle}\to 0$ as well. Alternatively, the 
systematic divergence of nearby initial conditions exhibited by chaotic orbits 
would suggest that, for an initially localised ensemble, quantities like 
${\sigma}_{px}(t)$ should {\it grow} exponentially at early times.
\vskip .1in
\par\noindent{{\bf 2.2 Convergence towards a near-invariant distribution} 
{\normalsize $f_{niv}$}}
\vskip .1in
\par\noindent
Consider first the evolution of the six reduced distributions $f(Z_{a},Z_{b})$
without any temporal averaging. Here one finds that, for ensembles of initial 
conditions corresponding to unconfined chaotic orbits, i.e., chaotic orbits 
not trapped by cantori near regular islands, there is a coarse-grained 
approach towards a near-invariant distribution $f_{niv}$ which, when 
quantified in terms of either an $L^{1}$ or $L^{2}$ norm, proceeds 
exponentially in time, i.e., $Df(t){\;}{\;}{\propto}{\;}{\exp}(-{\Lambda}t)$. 
At very early times, when $f$ has nonvanishing support only in a small number 
of cells, plots of the $L^{1}$- and $L^{2}$- distances can differ 
significantly, the $L^{2}$-distance starting larger but decreasing 
substantially more quickly. However, once the initial ensemble has spread out
so as to have nonzero support in a significant fraction of the cells, the 
$L^{1}$ and $L^{2}$ rates agree, at least approximately. This suggests that
the rate of convergence towards a near-invariant $f_{niv}$ is insensitive
to the detailed choice of norm, a result that was confirmed by exploring the
effects of allowing for other $L^{p}$ norms. In all cases, the rate 
${\Lambda}$ corresponds to a time scale ${\tau}{\;}{\equiv}{\;}{\Lambda}^{-1}$ 
which is comparable to, albeit somewhat longer than a characteristic crossing 
time $t_{cr}$. In other words, the approach towards a near-invariant $f_{niv}$ 
proceeds on the natural time scale $t_{cr}$ or, equivalently, the time scale 
${\chi}^{-1}$ on which small perturbations in initial conditions would be 
expected to grow. 

This behaviour is illustrated in Fig.~1. Here panel (a) exhibits the $L^{1}$ 
and $L^{2}$ distances $Df(x,y)$ for an ensemble of unconfined chaotic orbits 
with energy $E=1.0$ evolved in the dihedral potential. At very early times,
the remaining five distances, $f(x,p_{x})$, $f(x,p_{y})$, $f(y,p_{x})$, 
$f(y,p_{y})$ and $f(p_{x},p_{y})$, not shown, look different from both $f(x,y)$
and each other but, after a time $t{\;}{\sim}{\;}10$ they exhibit similar 
curvatures. In each case, $Df$ decreases exponentially until a time 
${\sim}{\;}40$, and then asymptotes towards a near-constant value. This 
later time behaviour is a finite size effect. Neither the near-invariant 
$f_{niv}$ nor the true time-dependent $f(t)$ is determined exactly, since each 
was generated from a finite number of orbits, so that, even if the computed 
$f(t)$ and $f_{niv}$ both sample the {\it true} ${\hat f}_{niv}$, the distance 
$Df$ between them will not vanish exactly. Fig. 1 (b) exhibits the 
corresponding $L^{1}$- and $L^{2}$-distances for a chaotic ensemble with 
$E=6.0$ again evolved in the dihedral potential. Once more the decay of all 
the $Df$'s is exponential, but the convergence rates ${\Lambda}_{ab}$ are 
substantially smaller, so that $Df$ continues to decrease until 
$t{\;}{\sim}{\;}100$. 

Initially localised ensembles of regular orbits, for which all the Lyapunov
exponents vanish identically, disperse much less efficiently. To the extent
that any approach towards a near-invariant $f_{niv}$ is observed, it will
proceed as a power law in time, and the characteristic time scale is typically
long compared with the natural crossing time $t_{cr}$. 

Ensembles corresponding to chaotic orbits originally confined near regular 
islands by cantori behave somewhat differently. Once again there is an 
approach towards a near-invariant $f_{niv}$ which, in many cases, is 
reasonably well fit by an exponential (although a power law fit often proves
nearly as good!). However the rates ${\Lambda}$ tend to be significantly 
smaller than the rates associated with unconfined ensembles with the same 
energy. This correlates with the fact that the largest short time Lyapunov
exponent (see, e.g., Grassberger, Badii, \& Politi 1988) for a confined 
chaotic segment,
albeit nonzero, is typically much smaller than the largest short time exponent
for an unconfined segment with the same energy (Mahon, Abernathy,
Bradley, \& Kandrup 1995). 
Figs.~1 (c) and (d) exhibit the distances $Df(x,p_{x}$ and $Df(y,p_{y})$ for 
an ensemble of confined chaotic orbits with $E=6.0$. Here decreases in the 
$Df$'s are reasonably well fit by an exponential, but it is clear that the 
rates ${\Lambda}_{ab}$ are much smaller than those associated with the 
ensemble from which Fig.~1 (b) derives.

The near-invariant $f_{niv}$ associated with an
ensemble of confined chaotic orbits is typically very different from the 
$f_{niv}$ associated with an unconfined ensemble, avoiding as it does large
portions of the stochastic sea far from any regular islands. If the initial
conditions be integrated for a sufficiently long time, often $t{\;}{\sim}{\;}
1000$ or more, the resulting orbits eventually diffuse through cantori to
probe these depths of the stochastic sea and approach a true invariant 
distribution. Ensembles of initially unconfined chaotic orbits also exhibit
diffusion through cantori on a comparable time scale as some fraction of the
orbits become trapped near regular islands. However, this is usually a much 
less conspicuous effect since, at least for the three potentials considered
here, the volume of the confined chaotic regions is typically much smaller
than the volume of the unconfined chaotic region.

When probing the initial approach towards a near-invariant 
$f_{niv}(Z_{a},Z_{b})$ exhibited by an ensemble of unconfined chaotic orbits, 
it is also useful to quantify the degree to which the best fit rate 
${\Lambda}_{ab}$ is approximately the same for all six choices of reduced 
distribution. In most cases, the answer appears to be: yes. If best fit slopes 
are extracted for all six possibilities and their values compared, one 
typically finds agreement at the $5-10\%$ level. However, one {\it does} see 
occasional exceptions where, compared with the remaining ${\Lambda}$'s, one of 
the ${\Lambda}_{ab}$'s is especially small and another is especially large. In 
virtually every case, the exceptional distributions are $f_{niv}(x,p_{x})$ and 
$f_{niv}(y,p_{y})$. Physically, an especially small ${\Lambda}_{xpx}$ and an 
especially large ${\Lambda}_{ypy}$ corresponds to a flow in which, when 
viewed in configuration space, initially proximate orbits tend to diverge much 
more quickly in the $x$-direction than in the $y$-direction, which implies in 
turn larger divergences in the $p_{x}$ than the $p_{y}$ direction. It follows 
that, respectively, $f(x,p_{x},t)$ and $f(y,p_{y},t)$ approach near-invariant
distributions somewhat faster and slower that the remaining four reduced
distributions which involve one fast and one slow variable. 
Figs.~1 (a) and(b) correspond to ensembles where all six rates are comparable. 
Fig.~1 (c) and (d) corresponds to an ensemble where ${\Lambda}_{xpx}$ is 
especially small and ${\Lambda}_{ypy}$ is especially large.

Another significant point is that, modulo the aforementioned examples of
especially fast and slow convergences, which depend rather sensitively on the
choice of ensemble, the convergence rates for different ensembles of chaotic
orbits with the same energy tend to be quite similar. Specifically, to the 
extent that orbit segments in different ensembles yield distributions 
of short time Lyapunov exponents that are very similar, they may be expected to
yield convergence rates that are nearly the same. However, small, but
statistically significant, differences between the distributions of short time
Lyapunov exponents can be observed for different chaotic ensembles, even two 
ensembles with the same energy both comprised entirely of unconfined chaotic 
orbits; and these differences, which manifest the fact that the detailed 
chaotic behaviour is slightly different, can translate into significant 
differences 
in the rates at which the ensembles evolve towards near-invariant $f_{niv}$'s. 

When comparing ensembles of unconfined chaotic orbits with different
energies, one also finds that, overall, the observed rates of convergence 
${\Lambda}$ correlate reasonably well with the mean value ${\overline\chi}$ of 
the largest short time Lyapunov exponent in the sense that increases in 
${\Lambda}$ coincide with increases in ${\overline\chi}$. However, the ratio 
${\cal R}={\Lambda}/{\langle}{\chi}{\rangle}$ does not always seem to assume
a constant value, as was found in many cases by Kandrup \& Mahon (1994) 
when considering reduced distributions ${\tilde f}(Z_{a},Z_{b})$ that 
averaged over several successive time steps.

Finally, it should be noted that, in addition to everything else, a plot of 
$Df(t)$ can exhibit considerable structure superimposed on the exponential 
approach, the details of which depend on (i) which reduced distribution one 
considers, (ii) which energy (and class of chaotic orbit) that one probes, and 
even (iii) which specific ensemble of initial conditions one chooses. The fact 
that, overall, $Df(t)$ decays to zero exponentially is a robust statement but,
as noted, e.g., by Merritt \& Valluri (1996), it does not constitute the whole 
story. An extreme example of this is exhibited in Fig.~1 (e) and (f), which 
exhibit two of the $Df$'s for an ensemble of unconfined chaotic orbits with 
$E=4.0$ evolved in the dihedral potential. For this and most other ensembles,
$Df(x,p_{x})$ and $Df(y,p_{y})$ exhibit relatively little structure 
superimposed on their exponential decrease, whereas the four remaining $Df$'s
exhibit larger amplitude oscillations. (Similar $Df(x,y)$'s can also be 
observed when one performs Langevin simulations, generating multiple 
realisations of single initial condition evolved in the presence of low 
amplitude fricton and noise, and tracks the evolved ensemble of orbits as it 
approaches a near-invariant distribution (see, e.g., Fig.~7 in Habib, Kandrup, 
\& Mahon 1997).

The results described hitherto all pertain to the behaviour of individual 
snapshots, without any temporal coarse-graining. However, 
it is also interesting to determine what happens if instead one considers a 
coarse-grained ${\tilde f}(Z_{a},Z_{b},t)$ which averages over several 
successive time steps. This was explored for the potentials (2) - (4) by 
analysing data recorded at various intervals between ${\Delta}t=0.25$ and 
${\Delta}t=2.0$, which were averaged over $k$ successive time steps. An 
investigation of the effects of increasing $k$ facilitated two concrete 
conclusions:
\par\noindent
1. As probed by the $L^{1}$ distance, averaging over several successive
snapshots leads generically to a significant increase in the rate at which
an ensemble evolves towards a near-invariant distribution. By contrast,
averaging has a comparatively minimal effect on the computed $L^{2}$ distance.
For this reason the $L^{1}$ and $L^{2}$ convergence rates ${\lambda}_{ab}$
associated with data averaged over successive snapshots are in better agreement
than the unaveraged rates ${\Lambda}_{ab}$, although the $L^{1}$ rates tend
to remain somewhat smaller. 
\par\noindent
2. The exact number of time steps over which one averages is relatively
unimportant. Assuming that snapshots are recorded at intervals
${\Delta}t{\;}{\sim}{\;}t_{cr}$, averages over different $k$'s extending from 
$5<k<25$ lead relatively similar results; and, similarly, varying ${\Delta}t$
from ${\sim}{\;}0.25t_{cr}$ to $2.0t_{cr}$ has a comparatively minor effect.

A concrete example of this behaviour is shown in Fig.~2, which plots the 
$L^{1}$ and $L^{2}$ distances $D{\tilde f}(x,y,t)$ computed for the same 
ensemble of unconfined chaotic orbits used to generate Fig.~1 (a). This Figure 
was generated by recording data at intervals ${\Delta}t=0.5$ and then 
averaging over $k=3$ and $15$ time steps, corresponding to intervals 
extending from ${\Delta}T=1.5$ to ${\Delta}T=7.5$. The specific
value associated with time $t$ involves an average from times $t$ to $t+0.5k$.
Overall, the $L^{1}$ slopes associated with $5<k<25$ are similar to one
another and somewhat larger in magnitude than that slopes for 
$1{\;}{\le}{\;}k{\;}{\le}{\;}3$ (the value $k=1$ corresponds to Fig.~1(a)). By
contrast, the $L^{2}$ slopes for$1{\;}{\le}{\;}k{\;}{\le}{\;}3$ do not differ 
substantially from the slopes computed with larger $k$. 

Overall, comparing different values $5{\;}{\le}{\;}k{\;}{\le}{\;}25$ for a 
single ensemble with the same reduced ${\tilde f}(Z_{a},Z_{b})$ yields 
agreement at the $4-8\%$ level. Comparing the mean ${\lambda}_{ab}$ for the 
same ensemble but different reduced distributions typically yields agreement 
at the $2-4\%$ level (except for the aforementioned exceptional cases where
one ${\Lambda}_{ab}$ is especially large and another especially small.. 
Comparing the typical values of 
${\Lambda}_{ab}$ and/or ${\lambda}_{ab}$ for different ensembles with the same
energy indicates that, when the mean short time ${\overline\chi}$'s are 
very similar, the computed ${\Lambda}_{ab}$ and/or ${\lambda}_{ab}$ will also
coincide. Comparing the typical values of ${\Lambda}_{ab}$ and ${\lambda}_{ab}$
for ensembles with different energies demonstrates that there is a reasonable
correlation between ${\overline\chi}$ and both ${\Lambda}_{ab}$ and 
${\lambda}_{ab}$. Energies for which unconfined chaotic orbits have larger
values of ${\overline\chi}$ tend systematically to have larger values of
${\Lambda}_{ab}$ and ${\lambda}_{ab}$ as well. 

The trends connecting ${\overline\chi}$, ${\Lambda}_{ab}$, and ${\lambda}_{ab}$
are exhibited in Fig.~3, which plot $0.5{\overline\chi}$ (solid curve), 
the $L^{2}$ rate ${\Lambda}_{L2}$ for individual snapshots (dashed curve),
the $L^{1}$ rate ${\Lambda}_{L1}$ for individual snapshots (dot dashed), and 
the $L^{1}$ rate ${\lambda}_{L1}$ generated by averaging over different 
coarse-grainings with $5<k<25$ (triple-dot-dashed). Each ${\Lambda}$ and 
${\lambda}$ involved averaging over all six possible reduced distributions. 
\vskip .1in
\par\noindent{\bf 2.3 Evolution of lower order moments}
\par\noindent
\vskip .1in
Turn now to the evolution of lower order moments, considering first 
quantities like ${\langle}x{\rangle}$ which, when computed using the invariant 
$F_{iv}(x,y,p_{x},p_{y})$ appropriate for an ensemble of chaotic orbits, 
vanish by symmetry. Here one finds that, for ensembles of initial conditions 
corresponding to unconfined chaotic orbits, such moments as 
${\langle}x{\rangle}$, ${\langle}xy{\rangle}$, or ${\langle}xy^{2}{\rangle}$ 
do indeed evolve towards zero, the expected result for a microcanonical 
sampling of the chaotic portions of the constant energy hypersurface. A plot 
of these decaying moments can exhibit a considerable amount of structure but, 
nevertheless, one finds that, in almost every case, the envelop 
approaches zero exponentially. 

An obvious question is whether, for a given ensemble, the moments which decay 
to zero exponentially all do so at the same, or at least comparable, rates. 
The answer here is that, overall, the rates are often very similar. Consider, 
e.g., the linear moments ${\langle}x{\rangle}$, ${\langle}y{\rangle}$, 
${\langle}p_{x}{\rangle}$, and ${\langle}p_{y}{\rangle}$ which, for the 
Hamiltonians considered in this paper, should vanish when evaluated for an 
$E=$constant $F_{iv}(x,y,p_{x},p_{y})$. For those energies where the phase 
space is relatively simple and obstructions like cantori play a relatively 
minimal role, the inferred rates for these moments are nearly identical, 
agreeing at the $5-10\%$ level. 
However, for energies where the phase space is more complicated the moments 
associated with one direction, either $x$ or $y$, can exhibit a much slower 
approach towards equilibrium, the rate ${\Gamma}$ being less than half as 
large. The ensembles where this behaviour is observed are precisely those for
which, as described in $\S$ 2.2, one $Df(Z_{a},Z_{b},t)$ decays especially
slowly and another especially fast.

Quadratic cross moments like ${\langle}xy{\rangle}$ or 
${\langle}xp_{y}{\rangle}$ and cubic moments like ${\langle}x^{3}{\rangle}$, 
${\langle}xy^{2}{\rangle}$ or ${\langle}xyp_{x}{\rangle}$ also tend to decay 
at comparable rates. However, there is a slight tendency, not manifested in all
cases, for some pairs e.g., ${\langle}xp_{x}{\rangle}$ or 
${\langle}yp_{y}{\rangle}$, to decay somewhat slower than the other quadratic
moments, which reflects the fact that correlations between a coordinate and
its conjugate momentum tend to decay comparatively slowly. In addition, plots 
of the nonlinear moments tend to exhibit more structure than do plots of the 
linear moments. 

This behaviour is illustrated in Fig.~4, which exhibits moments generated from
the same ensemble of orbits in the dihedral potential with $E=1.0$ which was
used to generate Fig.~1 (a). Panel (a) - (e) exhibit, respectively, one 
representative linear and four quadratic moments. Most of the moments decay at 
comparable rates, although ${\langle}xp_{x}{\rangle}$ and 
${\langle}yp_{y}{\rangle}$ (the latter not shown) clearly decay substantially 
more slowly than the other quadratic moments. 

Another obvious question is whether the rates associated with the decay towards
zero are the same for different orbit ensembles with the same energy.
Here the answer is that, overall, the rate (or largest rate, if several widely
disperate rates are observed) is comparatively insensitive to the choice of
ensemble, provided that one restricts attention to unconfined chaotic orbits.
This largest rate typically ranges between $1/3$ and $2/3$ the value of
${\overline\chi}$, the mean largest short time Lyapunov exponent associated 
with the orbit ensemble, which implies a value intermediate between the rates 
${\Lambda}_{L1}$ and ${\Lambda}_{L2}$ associated with the $L^{1}$ and $L^{2}$ 
convergence of $f(Z_{a},Z_{b},t)$ towards $f_{niv}$. 


Moments generated from ensembles of regular initial conditions evolve very
differently. For regular ensembles, the moments show no evidence of an 
exponential decay on a time scale ${\sim}{\;}{\chi}^{-1}$. In general, the 
moments will not decay towards zero and, even if they do, the decay is much 
better fit by a slow power law than by an exponential.

Now consider combinations of moments like the dispersion ${\sigma}_{x}=
\sqrt{{\langle}x^{2}{\rangle}-{\langle}x{\rangle}^{2}}$ which, for a localised 
ensemble of initial conditions, start small but eventually grow large.
For ensembles of unconfined chaotic orbits, such quantities typically grow 
exponentially in  time until they becomes ``macroscopic,'' so that, e.g., the 
magnitude of ${\sigma}_{x}$ becomes comparable to the linear size of the 
configuration space region accessible to orbits with the specified energy. 
Moreover, one usually finds that all four dispersions, ${\sigma}_{x}$, 
${\sigma}_{y}$, ${\sigma}_{px}$, and ${\sigma}_{py}$, grow at similar rates 
which are comparable to, but somewhat larger than, the mean short time 
exponent ${\overline\chi}$  for the ensemble. By contrast, when evaluated
for ensembles comprised of regular orbits, quantities like ${\sigma}_{x}$ grow
only much more slowly in a fashion reasonably well fit by a power law with 
an index $p$ of order unity, i.e., ${\sigma}_{x}{\;}{\propto}{\;}t^{p}$ with
$p{\;}{\sim}{\;}1.0$. Ensembles comprised of confined chaotic orbits typically
exhibit an intermediate behaviour. 
The typical behaviour of the dispersions for ensembles of unconfined chaotic 
orbits is illustrated in Fig.~5, which was again generated for the 
orbit ensemble with $E=1.0$ used to create Fig.~1. 

The aforementioned behaviour for the growing modes is easy to understand: If
the ensemble is chaotic, there is a systematic tendency for nearby trajectories
to diverge at a rate set by the largest Lyapunov exponent for individual 
orbits. That the growth rate for the dispersion is larger than 
${\overline\chi}$ reflects the obvious fact that this rate is dominated by 
those orbits 
that are most unstable. Differences between the growth rates in different 
directions can be attributed to the fact that, on the average, orbits in the 
ensemble will be exponentially unstable at somewhat different rates in 
different directions. If, alternatively, the orbits are regular, nearby 
trajectories only diverge as a power law in time, so that the dispersions 
exhibit a much slower power law growth. It should also be noted that the same 
qualitative behaviour arises when tracking the evolution of multiple 
realisations of the same initial condition integrated in the presence of low 
amplitude friction and noise (see, e.g., Habib, Kandrup, \& Mahon 1997). 
There once again the Lyapunov exponent sets the time scale (if any) on which 
different orbits in the ensemble diverge exponentially, the average rate of
divergence being somewhat larger than the mean ${\overline\chi}$.

Finally it may be noted that growing moments can be combined in interesting 
ways to yield quantities that decay exponentially in time. Thus e.g., one 
finds that 
$Dx{\;}{\equiv}{\;}{\langle}x^{4}{\rangle}-2{\langle}x^{2}{\rangle}^{2}$ and
$Dy{\;}{\equiv}{\;}{\langle}x^{4}{\rangle}-2{\langle}y^{2}{\rangle}^{2}$ both 
decay towards values that are very close to zero. This exponential decay 
is illustrated in Fig.~4 (f), which exhibits ${\rm ln}\,Dx$ for the orbit 
ensemble used to generate Fig.~1 (a).
\vskip .15in
\par\noindent
{\bf 3 THREE-DIMENSIONAL SYSTEMS}
\vskip .1in
\par\noindent
{\bf 3.1 Description of what was computed}
\vskip .1in
\par\noindent
The conclusions described in this Section derive from a study of Hamiltonian
systems of the form 
$$H={1\over 2}{\Bigl(}p_{x}^{2}+p_{y}^{2}+p_{z}^{2}{\Bigr)}+V(x,y,z),\eqno(6)$$
with 
$$V(x,y,z)=-(x^{2}+y^{2}+z^{2})+{1\over 4}(x^{2}+y^{2}+z^{2})^{2}
-{1\over 4}(x^{2}y^{2}+ay^{2}z^{2}+bz^{2}x^{2}), \eqno(7) $$
where $a$ and $b$ are real-valued constants. The most symmetric choice, 
$a=b=1$, yields the obvious three-dimensional analogue of the two-dimensional 
dihedral potential, which now manifests cubic, rather than square, symmetry. 
Selecting $a$ or $b{\;}{\ne}{\;}1$ breaks this symmetry. For generic values 
of $a$ and $b$, at most energies the phase space admits a coexistence of 
significant measures of both regular and chaotic orbits, much as for the 
two-dimensional dihedral potential. However, adding a third dimension seems 
to make the chaotic orbits somewhat less unstable in the sense that, at least
for $a=b=1$, the largest Lyapunov exponent for the three-dimensional system 
tends to be somewhat smaller than the largest exponent for two-dimensional 
systems (at least for relatively low energies). 
The special case $a=b=2$ leads to the partially integrable potential
$$V(x,y)=-(x^{2}+y^{2})+{1\over 4}(x^{2}+y^{2})^{2}-{1\over 4}x^{2}y^{2} 
+{\Bigl(}{1\over 4}z^{2}-z^{2}{\Bigr)} , \eqno(8) $$
for which motion in the $z$-direction is decoupled from motion in the 
remaining $x$- and $y$-directions.

Ensembles of $6561$ initial conditions were chosen in two different ways. 
In some cases, these were obtained by uniformly sampling a square in
the $y-p_{y}$ (or $z-p_{z}$) plane, setting $x=0$, selecting fixed values for 
$z$ and $p_{z}$ (or $y$ and $p_{y}$),
and then solving for $p_{x}>0$. In other cases, they were obtained by uniformly
sampling a four-dimensional $y$-$z$-$p_{y}$-$p_{z}$ hypercube with sides
${\Delta}y={\Delta}z={\Delta}p_{y}={\Delta}p_{z}=0.2$, setting $x=0$, 
and solving for $p_{x}>0$. The resulting orbits were analysed much as for the
two-dimensional systems described in Section 2, except that attention focused
on all $15$ reduced distributions $f(Z_{a},Z_{b})$ and all possible moments
${\langle}x^{i}y^{j}z^{k}p_{x}^{l}p_{y}^{m}p_{z}^{n}{\rangle}$ with 
$i+j+k+l+m+n{\;}{\le}{\;}4$. One major difference observed for three-, as
opposed to, two-dimensional systems was that, even for $a=b=1$, the time 
required for a chaotic ensemble to approach a near-invariant $f_{niv}$ could 
be extremely long, so that obtaining a reasonable approximation to $f_{niv}$ 
sometimes required integrating an orbital ensemble for times as large as 
$t{\;}{\sim}{\;}1000$. In most cases, only the largest short time Lyapunov 
exponent was computed. However, in some cases a more complicated code (Wolf, 
Swift, Swinney, and Vastano 1985) was used to extract the full spectrum of 
exponents.
\vskip .1in
\par\noindent
{\bf 3.2 Convergence towards a near-invariant distribution} 
{\normalsize $f_{niv}$}
\vskip .1in
\par\noindent
Overall, the behaviour exhibited by localised ensembles of initial conditions
evolved in three-dimensional potentials is quite similar to that observed
in two-dimensional potentials. Ensembles of chaotic orbits, not initially 
stuck near a regular island, typically exhibit an exponential approach towards 
a near-invariant distribution $f_{niv}$ at a rate ${\Lambda}$ which is 
comparable in magnitude to ${\overline\chi}$, the mean value of the largest 
short time Lyapunov exponent. Alternatively, ensembles of regular orbits 
exhibit little, if any, tendency to evolve towards a near-invariant $f_{niv}$.

However, there {\it are} some striking differences. 
One important difference between two- and three-dimensional systems is that,
for the latter, the approach towards a near-invariant $f_{niv}$ often proceeds
at grossly different rates in different directions. Here the determining
factor is {\it not} in general the precise values of $a$ and $b$ (provided one 
avoids the special case $a=b=2$, which leads to a partially integrable 
system). Rather, what seems to matter more is the location of the phase space
cell that was chosen to generate the initial conditions. If, e.g., one chooses
an ensemble of initial conditions with relatively small values of $|z|$ and
$|p_{z}|$, this corresponding to orbits originally moving very nearly in the
$x-y$ plane, one often finds that the approach towards a near-invariant 
distribution
in the $z$ and $p_{z}$ directions is extremely slow. This, however, does not
imply that, for such ensembles, the largest short time Lyapunov exponent is
especially small. Indeed, if the initial $|z|$ and $|p_{z}|$ are both 
sufficiently small, the largest short time ${\overline\chi}$ for a 
three-dimensional system with $a=b=1$ will more closely approximate the 
largest Lyapunov exponent ${\chi}$ for a two-dimensional system than the 
largest ${\chi}$ for a three-dimensional system. 

Such a direction-dependent approach towards a near-invariant $f_{niv}$ is
illustrated in Figs.~6 (a), (c), and (e). These panels, generated from a 
chaotic ensemble with energy $E=6.0$ and an initial $z=0.2$ and $p_{z}=0.0$, 
evolved in the potential (7) with $a=b=1$, exhibits three different $L^{1}$ 
and $L^{2}$ distances, namely $Df(x,y)$, $Df(z,p_{x})$, and $Df(z,p_{z})$.
For all fifteen reduced distributions $Df(Z_{a},Z_{b})$, the approach 
towards a near-invariant $f_{niv}$ is exponential in time, but three 
significantly different rates are observed. The six $f(t)$'s involving only 
the ``fast'' variables $x$, $y$, $p_{x}$, and $p_{y}$ approach $f_{niv}$ 
especially quickly, whereas $f(z,p_{z},t)$, which involves the ``slow'' 
variables $z$ and $p_{z}$, evolves towards $f_{niv}$ especially slowly. The 
remaining eight $f(t)$'s, which involve one fast and one slow variable, 
approach $f_{niv}$ at an intermediate rate.

Not surprisingly, this direction-dependence becomes even more striking for
chaotic ensembles evolved with $a=b=2$, where motion in the 
$z$-direction is integrable. In this case, one finds that, irrespective of
the qualitative character of motion in the remaining four phase space 
directions, there is no strong exponential approach towards a near-invariant
distribution associated with either the $z$ or the $p_{z}$ direction. For 
chaotic ensembles corresponding to orbits with one positive Lyapunov exponent, 
one finds that the six reduced distributions involving $x$, $y$, $p_{x}$, and 
$p_{y}$ all exhibit an exponential approach towards a near-invariant $f_{niv}$,
typically at a rate comparable to what was observed for the two-dimensional
dihedral potential. However, the remaining nine reduced distributions,
all of which probe the $z$ and/or $p_{z}$ direction, exhibit a much weaker
tendency to approach a near-invariant $f_{niv}$. The quantities
$Df(Z_{a},Z_{b},t)$ with $Z_{a}=x,y,p_{x},$ or $p_{y}$ and $Z_{b}=z$ or $p_{z}$
{\it do} exhibit modest decreases, but $Df(z,p_{z},t)$ often shows essentially 
no tendency to decrease.

This behaviour is illustrated in Figs.~6 (b), (d), and (f), which exhibits the 
same $L^{1}$ and $L^{2}$ distances as did Fig.~6 (a), (c), and (e), again 
generated for an ensemble of chaotic initial conditions with $E=6.0$ but now 
evolved with $a=b=2$. These panels should be constrasted with Figs.~7 (a) and
(b), which exhibit two representative $Df$'s computed for an ensemble of
regular orbits evolved with $a=b=1$. It is apparent that, as would be expected
for this latter ensemble, these distances do not exhibit an efficient 
exponential damping, although the $L^{2}$ distances can exhibit a slow, albeit 
systematic, decrease.

Modulo the partially integrable case $a=b=2$ and special initial conditions 
like those initially stuck near the $x-y$ plane, one finds typically that, as 
for the case of two-dimensional potentials, the convergence rates 
${\Lambda}_{ab}$ associated with different reduced distributions
$f(Z_{a},Z_{b},t)$ computed for the same ensemble of chaotic orbits tend to be 
comparatively similar. However, a comparison of results from different 
ensembles exhibits more diversity than what was observed in two-dimensional
systems. Indeed, when sampling different portions of a single connected phase 
space region, all corresponding to chaotic orbits initially located relatively 
far from any large regular islands, one often finds convergence rates 
${\Lambda}_{ab}$ that differ by substantially more than $10\%$. This diversity
appears to correlate with the fact that different ensembles extracted from the 
same connected phase space region can be characterised by distributions of 
short time Lyapunov exponents, $N[{\chi}]$, which differ substantially more 
than what was observed for the corresponding distributions in 
two-dimensional systems. 

It is also interesting to determine the effects of averaging over successive
snapshots. For ensembles where the orbits behave chaotically in all three
spatial dimensions, averaging over two or more successive snapshots has much 
the same effect as for two-dimensional systems. In particular, the $L^{1}$ 
rate of convergence increases if one averages over two or
more snapshots and, when averaged over (say) five or more snapshots, appears
to asymptote towards a value ${\lambda}_{ab}$ that is insensitive to the 
precise number $k$ of successive time steps used to compute the average. When
considering orbit ensembles that are chaotic in two directions but regular in 
the third, one also finds that, at least in the chaotic directions, averaging
yields the same behaviour as for two-dimensional systems. However, when 
probing the integrable direction (if any) associated with a flow, temporal 
averaging has a different effect. In this case, averaging over successive time 
steps {\it can} amplify at least slightly a very weak tendency to evolve 
towards some $f_{niv}$. More important, however, is the fact that the 
averaging {\it per se} yields a ${\tilde f}$ that more closely approximates 
$f_{niv}$, even at very early times. The reason for this is not hard to see. 
Because motion in the $z-p_{z}$ plane is periodic, it is easy to associate a 
near-invariant $f_{niv}(z,p_{z})$ with even a single orbit by computing the 
relative amount of time that the orbit spends in the neighbourhood of each 
point in the $z-p_{z}$ plane. The obvious point then is that if a localised 
ensemble of initial conditions exhibits no appreciable tendency to disperse, 
averaging over a sufficiently large number of successive snapshots will yield 
a decent approximation to $f_{niv}$ which, for a fixed number of equally 
spaced snapshots, should not differ appreciably when evaluated for early and
late intervals intervals, $t_{1}<t<t_{2}$ and $t_{1}+T<t<t_{2}+T$.
\vskip .1in
\par\noindent
{\bf 3.3 Evolution of lower order moments}
\vskip .1in
\par\noindent
An examination of moments generated from ensembles evolved in three-dimensional
potentials leads to two general conclusions:
\par\noindent
1. When considering ensembles of orbits that are wildly chaotic in all three
directions, one finds the same qualitative behaviour as was observed for
wildly chaotic orbits evolved in two-dimensional systems. Moreover, if an
orbit is wildly chaotic in two directions, but regular or much less chaotic
in the third direction, one finds that moments probing the two wildly chaotic
directions again resemble the moments derived from wildly chaotic ensembles 
in two-dimensional systems.
\par\noindent
2. Just as the case of two-dimensional systems, there is a strong correlation
between the reduced distribution $f(Z_{z},Z_{b},t)$ and various moments. 
Suppose, e.g., that for some ensemble with two positive Lyapunov exponents,
one of the reduced distributions, say $f(z,p_{z},t)$ approaches an invariant
distribution much more slowly than the remaining distributions. One then finds
invariably that ${\langle}z{\rangle}$ and ${\langle}p_{z}{\rangle}$ evolve 
towards zero more slowly than the remaining first moments, and that 
${\sigma}_{z}$ and ${\sigma}_{pz}$ grow more slowly than the remaining 
dispersions. 

Aside from these trends, three dimensions leads to several new features which
reflect the possibility that the orbits can be integrable, or much less
chaotic, in one configuration space direction than in the remaining two
directions. The diversity of what can arise may be understood by contrasting
the behaviour of ensembles comprised of orbits that are wildly chaotic in all
direction, which yield moments closely resembling what is observed for 
unconfined chaotic orbits in two-dimensional systems, with three 
representative ensembles {\it not} wildly chaotic in all three directions, 
namely (1) an ensemble of completely regular orbits, (2) an ensemble of orbits 
which is integrable in the $z$-direction but wildly chaotic in the $x$- and 
$y$-directions, and (3) an ensemble where motion in all three directions is 
chaotic but the $z$-direction is much less chaotic than the $x$- and 
$y$-directions.

Figure 8, generated from the same regular ensemble used to create Fig.~7, 
exhibits four representative moments, namely ${\sigma}_{z}$, 
${\langle}z{\rangle}$, ${\langle}zp_{y}{\rangle}$, and 
${\langle}y^{2}z{\rangle}$. For this and other regular ensembles, one discovers
that, unlike the case in chaotic ensembles, the dispersions in all six phase 
space variables exhibit an overall growth which is linear, rather than 
exponential, in time. However, as is evident from Fig.~7 (a), this is not the 
whole story. The evolution of the dispersions really involves large amplitude 
oscillations characterised by an envelop that grows linearly in time. This 
sort of qualitative behaviour is also observed for other growing quantities 
such as ${\langle}y^{4}{\rangle}-{\langle}y{\rangle}^{4}$. Indeed, one finds 
that all the combinations of moments which grow exponentially for a chaotic 
ensemble exhibit a much weaker, oscillating power law growth when computed for 
a regular ensemble. It would appear that an initially localised ensemble of 
regular orbits cannot disperse exponentially.

Analogously, one finds that the six linear moments, and all other moments 
which, for a chaotic ensemble, approach zero exponentially, exhibit only a
modest tendency to damp. Indeed, to the extent that these moments decrease at 
all, that decrease is better fit by a power law than an exponential. The other 
obvious point is that, like the dispersions, these moments exhibit coherent 
oscillations. Indeed, overlaying plots of (say) ${\sigma}_{z}$ and 
${\langle}z{\rangle}$ makes it clear that these quantities oscillate with 
essentially the same frequency, a fact that reflects the periodicity of the 
regular orbits in the ensemble. Finally, it should be noted that, in this case,
not all the cross moments evolve towards zero. For example, the coherence 
manifested by the orbits in the ensemble would seem to imply that 
${\langle}zp_{y}{\rangle}$ is evolving towards a value 
${\langle}zp_{y}{\rangle}{\;}{\approx}{\,}-2.9$ rather than zero.

Figure 9 exhibits the same four moments, now computed for the wildly chaotic
ensemble used to generate Figs.~6 (b), (d), and (f), with $a=b=2$, so that
motion in the $z$ direction is integrable. For this ensemble, the moments 
involving only $x$, $y$, $p_{x}$, and $p_{y}$, not shown here, all behave 
qualitatively like moments computed for unconfined chaotic orbits in 
two-dimensional potentials, such as is exhibited in Fig.~1 (a) and (b).
Alternatively, as is illustrated in Fig.~9 (a) and (b), moments involving 
only $z$ and/or $p_{z}$ behave qualitatively like moments associated with 
completely regular orbits. 

In some cases, moments involving both 
$z$ and/or $p_{z}$ {\it and} $x$, $y$, $p_{x}$, and/or $p_{y}$ resemble 
moments for purely chaotic two-dimensional orbits, but in other cases they
resemble results appropriate for regular orbits. That the cross moment 
${\langle}y^{2}z{\rangle}$ behaves very much like the regular moment 
${\langle}z{\rangle}$ is easily understood. Because motion in the $y$-direction
is wildly chaotic, except for very early on, there is little, if any, 
correlation between the values of $y$ and $z$ assumed by individual orbits, so 
that
${\langle}y^{2}z{\rangle}{\;}{\approx}{\;}{\langle}y^{2}{\rangle}\,{\langle}
z{\rangle}$. However, the orbit ensemble disperses sufficiently fast in the
$y$-direction that, within a time $t{\;}{\sim}{\;}15-20$, 
${\langle}y^{2}{\rangle}$ has asymptoted towards a constant value, so
that ${\langle}y^{2}z{\rangle}{\;}{\propto}{\;}{\langle}z{\rangle}$.
Similarly, one would expect that ${\langle}zp_{y}{\rangle}{\;}{\approx}{\;}
{\langle}z{\rangle}{\langle}p_{y}{\rangle}$, which implies, in agreement with
what is actually found, that this moment should die to zero exponentially 
like ${\langle}p_{y}{\rangle}$.

Figure 10 exhibits once again the same four moments, now computed for the
ensemble used to construct Fig.~6 (a), (c), and (e), which was so chosen that, 
at least early on, motion in the $z$-direction is much less chaotic than in 
the $x$- and $y$-directions. Once more one finds that the moments involving 
only $x$, $y$, $p_{x}$, and $p_{y}$ behave qualitatively like moments for 
unconfined chaotic orbits in two-dimensional potentials. The dispersions all 
grow exponentially, at least initially, and the linear moments decay to zero 
exponentially. The linear moments ${\langle}z{\rangle}$ and 
${\langle}p_{z}{\rangle}$ also decrease in a fashion that is (at least) 
consistent with an exponential fit, although plots of these quantities show 
considerably more structure than what is observed for the other linear moments.

Very early on, ${\sigma}_{z}$ and ${\sigma}_{pz}$ 
appear to increase exponentially but, after a time $t{\;}{\sim}{\;}15$ or so,
the evolution of these dispersions is better fit by a power law. Indeed, 
the growth of ${\sigma}_{z}$ is so slow that one might naively conjecture
that motion is regular in the $z$-direction. In point of fact, however, this
is not so! A computation of the full spectrum of Lyapunov exponents indicates
that there are two positive Lyapunov exponents; and, even more blatantly, if 
one evolves the orbits for a sufficiently long time (in many cases $t>1000$
or more), one finds that they eventually begin to act wildly chaotic in all 
three directions. The fact that motion in the $z$-direction is chaotic,
albeit only moderately so, implies an absence of periodicity. This accounts
for the fact that Fig.~10 (a) shows less evidence for systematic oscillations
than do Figs.~8 (a) and 9 (a), as well as the fact that a plot of 
${\langle}zp_{z}{\rangle}$ for this ensemble is more irregular than 
correspomding plots for the ensembles used to generate Figs.~8 and 9. 
For the reason discussed in connection with Fig.~9, plots of moments like 
${\langle}y^{2}z{\rangle}$ again resemble plots of ${\langle}z{\rangle}$ and 
plots of ${\langle}zp_{y}{\rangle}$ again resemble plots of 
${\langle}p_{y}{\rangle}$.
\vskip .2in
\par\noindent
{\bf 4 CONCLUSIONS AND IMPLICATIONS}
\vskip .1in
\par\noindent
The principal implication of the work described here is that, in terms of
their approach towards an invariant, or near-invariant, distribution, regular
and chaotic flows behave very differently. For localised ensembles of initial 
conditions, chaotic flows exhibit an initial exponential divergence,
so that quantities like the dispersions in position and momentum increase
exponentially in time. Similarly, as probed by both (i) lower order moments
and (ii) coarse-grained reduced distribution functions, the ensemble will
exhibit an approach towards a near-invariant distribution which proceeds
exponentially in time. The time scale associated with both the initial 
divergence and the ultimate approach towards a near-invariant distribution is 
comparable to a characteristic crossing time $t_{cr}$, which in turn is
comparable to ${\chi}^{-1}$, where ${\chi}$ represents a typical positive
Lyapunov exponent. By contrast, regular ensembles exhibit an initial power 
law divergence and any approach towards a near-invariant distribution is 
typically better fit by a power law than an exponential. 

In chaotic two-dimensional systems, one finds that, for a fixed potential
and energy, different ensembles of unconfined chaotic orbits located far from 
any regular island tend to behave relatively similarly, although ensembles of
confined chaotic orbits, stuck near regular islands by cantori, can behave 
rather 
differently. Moreover, the approach towards a near-invariant distribution 
typically proceeds at comparable rates in different directions. In many cases, 
this behaviour persists in three-dimensional systems. However, this is not 
always true. If the potential is integrable in one direction, so that chaotic 
orbits have only one positive Lyapunov exponent, that direction will of course 
behave very differently from the other two directions. However, even for a 
fully nonintegrable system where chaotic orbits have two positive Lyapunov 
exponents the approach towards a near-invariant distribution can proceed at 
grossly different rates in different directions. 

The implication of all of this, in agreement with Merritt \& Valluri (1996), 
is that what they call chaotic phase mixing, the process whereby an ensemble 
of chaotic 
initial conditions loses its initial coherence and approaches a near-invariant 
distribution, is far more efficient than regular phase mixing. If the flow 
were completely chaotic and especially if, as probed by short time Lyapunov 
exponents, the degree of chaos associated with different parts of an orbit 
exhibited relatively little variability, the approach towards an invariant, or 
near-invariant, distribution would be relatively simple to visualise and 
understand. However, generic potentials admit a coexistence of both regular 
and chaotic orbits; and, for such potentials, different parts of the same 
chaotic orbit can manifest grossly different amounts of chaos. One thus 
anticipates that, even though chaotic phase mixing will always be more 
efficient than regular phase mixing, it could prove a relatively complex 
phenomenon that manifests a significant dependence on initial conditions.

Finally it is worth noting two important caveats.
\par\noindent
1. The operative feature that allows one to distinguish between ``chaotic'' and
``regular'' phase mixing is whether or not nearby orbits tend systematically
to diverge. In principle, however, a local divergence can obtain even if the
orbits are regular in the sense that, in an asymptotic $t\to\infty$ limit,
there are no positive Lyapunov exponents. All that is required is that the
``local stretching number,'' i.e., ``short time Lyapunov exponent,'' be 
non-zero.
\par\noindent
2. The fact that such quantities as ${\sigma}_{x}$ have asymptoted towards
a constant nonzero value, or that moments like ${\langle}p_{x}{\rangle}$
are nearly zero, does not imply the absence of any interesting subsequent
evolution. All that one really expects is that any subsequent evolution will
only proceed on relatively short scales. In general, one would anticipate
that, for an exponentially diverging flow, an initially localised ensemble 
will exhibit a bulk approach towards an invariant distribution, followed by
a more complicated evolution as power cascades down to progressively smaller
scales, in agreement with Fig.~2 in Lynden-Bell's (1967) original paper. A
concrete example of this behaviour is exhibited in Figs.~11. The four panels
here exhibit at times $t=2$, $4$, $8$, and $16$ the $x$ and $y$ coordinates of 
each of $225/times 225=50625$ unconfined chaotic orbits in the two-dimensional 
dihedral potential with $E=1.0$ which were generated by sampling the same 
phase space region as the ensemble used to generate Fig. 1 (a).

But what, if anything, might these conclusions imply about violent relaxation?
At least crudely, one might hope to visualise an evolution described by the 
collisionless
Boltzmann equation as involving a collection of characteristics corresponding
to orbits evolved in a specified time-dependent potential, ignoring the fact
that that potential is generated self-consistently. However, to the extent
that this picture is valid, one might anticipate that the efficacy with
which an initial ensemble approaches an equilibrium or near-equilibrium, i.e., 
a time-independent, or nearly time-independent, solution to the Boltzmann
equation, will depend on the degree to which the flow in the specified
potential is chaotic. In particular, to the extent that the flow is chaotic, 
so that many/most of the orbits are characterised by positive short time 
Lyapunov exponents, one would expect a rapid and efficient approach towards a 
near-equilibrium. 

Many galactic dynamicists have the intuition that ``realistic'' equilibrium
solutions to the collisionless Boltzmann equation correspond to integrable
or near-integrable potentials, which admit few if any chaotic orbits. However,
even if this be true, there would seem good reason to anticipate that the
time-dependent potential associated with the approach towards equilibrium 
will, at least initially, admit significant measures of chaotic orbits with 
positive short time Lyapunov exponents, 
Even assuming a high degree of spatial symmetry, a generic time-dependent 
potential $V({\bf r},t)$ will exhibit a substantial amount of chaotic behavour 
so that, at least until the time-dependence of $V$ becomes very weak and the
system is close to equilibrium, one would expect to see a significant amount 
of chaotic phase mixing which proceeds exponentially in time.

\vskip .2in
\par\noindent
{\bf ACKNOWLEDGMENTS}
\vskip .1in
\par\noindent
This work was supported in part by Los Alamos National Laboratory through the
Institute of Geophysics and Planetary Physics. Some of the numerical 
calculations described here were facilitated by computer time provided by 
{\it IBM} through the Northeast Regional Data Center (Florida). I am grateful
to Christos Siopis, Ilya Pogorelov, Donald Lynden-Bell, and Salman Habib 
for useful comments and interactions.
\vfill\eject
\par\noindent
Armbruster, D., Guckenheimer, J., Kim, S. 1989. Phys. Lett. A 140, 416.%
\par\noindent
Arnold, V. I. 1964, Russ. Math. Surveys 18, 85. %
\par\noindent
Aubry, S., Andre, G. 1978. in Bishop, A. R., Schneider, T. eds. Solitons and 
Condensed \par Matter Physics. Springer, Berlin, 264. %
\par\noindent
Binney, J., Tremaine, S. 1987. Galactic Dynamics. Princeton Univ.
Press, Princeton. 
\par\noindent
\par\noindent
Chandrasekhar, S. 1943. Principle of Stellar Dynamics. Univ. of Chicago
Press. Chicago.
\par\noindent
\par\noindent
Contopoulos, G. 1971. AJ 76, 147. %
\par\noindent
Grassberger, P., Baddi, R., Politi, A. 1988, J. Stat. Phys. 51, 135.
\par\noindent
Habib, S., Kandrup, H. E., Mahon, M. E. 1996, Phys. Rev. E 53, 5473.
\par\noindent
Habib, S., Kandrup, H. E., Mahon, M. E. 1997, ApJ 480, 155. 
\par\noindent
Habib, S., Kandrup, H. E., Pogorelov, I. V., Ryne, R. 1998, in preparation.
\par\noindent
Kandrup, H. E. 1998, ApJ 500, 120..
\par\noindent
Kandrup, H. E. 1998, Ann. N. Y. Acad. Sci, in press.
\par\noindent
Kandrup, H. E., Mahon, M. E. 1994, Phys. Rev. E 49, 3735.
\par\noindent
Lichtenberg, A. J., Lieberman, M. A. 1992, Regular and Chaotic Dynamics.
Springer, \par Berlin.
\par\noindent
Lynden-Bell, D. 1967, MNRAS 136, 101.
\par\noindent
MacKay, R. S., Meiss, J. D., Percival, I. C. 1984, Phys. Rev. Lett. 
52, 697. 
\par\noindent
Mahon, M. E., Abernathy, R. A., Bradley, B. O., Kandrup, H. E. 1995.
MNRAS 275, 443. %
\par\noindent
Mather, J. N. 1982, Topology 21, 45. %
\par\noindent
Merritt, D., Valluri, M. 1996. ApJ 471, 82.
\par\noindent
Morrison, P. J. 1980, Phys. Lett. A 80, 383.
\par\noindent
Pogorelov, I., Kandrup, H. E. 1998, in preparation.
\par\noindent
Toda, M. 1967. J. Phys. Soc. Japan 22, 431.
\par\noindent
Wolf, A., Swift, J. Swinney, H., Vastano, J. 1985, Physica D 16, 285.%
\vfill\eject
\centerline{\bf FIGURE CAPTIONS}
\par\noindent
Figure 1. (a) The $L^{2}$ (uppoer boldface curve) and $L^{1}$ (lower curve)
distance $Df(x,y,t)$ between $f(x,y,t)$ and a near-invariant $f_{niv}(x,y)$
computed for an ensemble of $6561$ unconfined orbits with $E=1.0$ evolved in
the dihedral potential (4). (b) The same for an ensemble of $6561$ unconfined 
chaotic orbits with $E=6.0$. (c) $Df(xp_{x},t)$ for a different ensemble of 
$6561$ unconfined chaotic orbits with $E=6.0$. (d) $Df(y,p_{y},t)$ for the
ensemble used to generate (c). (e) $Df(x,y,t)$ for an ensemble of $6561$ orbits
with $E=4.0$. (f) $Df(y,p_{y},t)$ for the ensemble used to generate (e).
\par\noindent
Figure 2 . The $L^{2}$- (upper boldface curve) and $L^{1}$- (lower curve)
distances $D{\tilde f}(x,y,t)$ for the ensemble used to construct Fig.~1(a), 
now considering a coarse-grained ${\tilde f}(x,y,t)$ which averages over a 
variable number $k$ of successive time steps separated by ${\Delta}t=0.5$. 
(a) $k=3$. (b) $k=15$. 
\par\noindent
Figure 3. The mean $L^{1}$ convergence rate ${\Lambda}_{L1}$ for individual 
snapshots (dot-dashed curve), the mean $L^{2}$ convergence rate 
${\Lambda}_{L2}$ averaging over between five and 25 time steps (triple-dot-
dashed curve), the mean $L^{2}$ convergence rate for individual snapshots
(dashed curve), and $0.5{\overline \chi}$ (solid curve), with 
${\overline\chi}$ the mean
value of the largest Laypunov exponent, for ensembles of unconfined chaotic
orbits with variable energy $E$. The error bars were generated by averaging 
over several different ensembles at each energy.
\par\noindent
Figure 4. Logarithmic plots of several moments computed for the ensemble of
unconfined chaotic orbits with $E=1.0$ evolved in the dihedral potential used 
to generate Fig.~1. (a) ${\rm ln}\,|{\langle}x{\rangle}|$.
(b) ${\rm ln}\,|{\langle}xy{\rangle}|$.
(c) ${\rm ln}\,|{\langle}p_{x}p_{y}{\rangle}|$.
(d) ${\rm ln}\,|{\langle}xp_{x}{\rangle}|$.
(e) ${\rm ln}\,|{\langle}xp_{y}{\rangle}|$.
(f) ${\rm ln}\,({\langle}x^{4}{\rangle}-2{\langle}x^{2}{\rangle}^{2})$
\par\noindent
Figure 5.  Logarithmic plots of dispersions computed for the ensemble used to
generate Fig.~7. (a) ${\rm ln}\,{\sigma}_{x}$. (b) ${\rm ln}\,{\sigma}_{px}$. 
\par\noindent
Figure 6. $L^{2}$ (upper boldface curve) and $L^{1}$ (lower curve) distances 
$Df(t)$ computed for an ensemble of $6561$ unconfined orbits with $E=6.0$ 
evolved in the three-dimensional potential with (i) $a=b=1$ and (ii) $a=b=2$.
(a) $Df(x,y,t)$ for the ensemble with $a=b=1$.
(b) $Df(x,y,t)$ for the ensemble with $a=b=2$.
(c) $Df(z,p_{x},t)$ for the ensemble with $a=b=1$.
(d) $Df(z,p_{x},t)$ for the ensemble with $a=b=2$.
(e) $Df(z,p_{z},t)$ for the ensemble with $a=b=1$.
(f) $Df(z,p_{z},t)$ for the ensemble with $a=b=2$.
\par\noindent
Figure 7. $L^{2}$ (upper boldface curve) and $L^{1}$ (lower curve) distances 
$Df(t)$ computed for an ensemble of $6561$ regular orbits with $E=4.0$ evolved 
in the three-dimensional potential (7) with $a=b=1$.
(a) $Df(x,y,t)$. (b) $Df(x,p_{y},t)$. 
\par\noindent
Figure 8. Four different moments computed for an ensemble of $6561$ regular
orbits with $E=4.0$, evolved in the three-dimensional potential (7) with
$a=b=1$. (a) The dispersion ${\sigma}_{z}$. (b). ${\langle}z{\rangle}$.
(c) ${\langle}zp_{y}{\rangle}$. (d) ${\langle}y^{2}{z}{\rangle}$.
\par\noindent
Figure 9. Four different moments computed for an ensemble of $6561$ chaotic
orbits with $E=6.0$, evolved in the three-dimensional potential (7) with
$a=b=2$. (a) The dispersion ${\sigma}_{z}$. (b). ${\langle}z{\rangle}$.
(c) ${\langle}zp_{y}{\rangle}$. (d) ${\langle}y^{2}{z}{\rangle}$.
\par\noindent
Figure 10. Four different moments computed for an ensemble of $6561$ chaotic
orbits with $E=6.0$, evolved in the three-dimensional potential (7) with
$a=b=1$. Initial conditions were chosen so that the orbit are initially
much less chaotic in the $z$-direction than the $x$- and $y$-directions.
(a) The dispersion ${\sigma}_{z}$. (b). ${\langle}z{\rangle}$.
(c) ${\langle}zp_{y}{\rangle}$. (d) ${\langle}y^{2}{z}{\rangle}$.
\par\noindent
Figure 11. The $x$ and $y$ coordinates of $50625$ unconfined chaotic orbits
evolved in the two-dimensional dihedral potential with $E=1.0$, generated from
initial conditions that sampled the same phase space region as did the ensemble
used to produce Fig.~1 (a). (a) Time $t=2$. (b) $t=4$. (c) $t=8$. (d) $t=16$.
\par\noindent
\vfill\eject
\pagestyle{empty}
\begin{figure}[t]
\centering
\centerline{
        \epsfxsize=12cm
        \epsffile{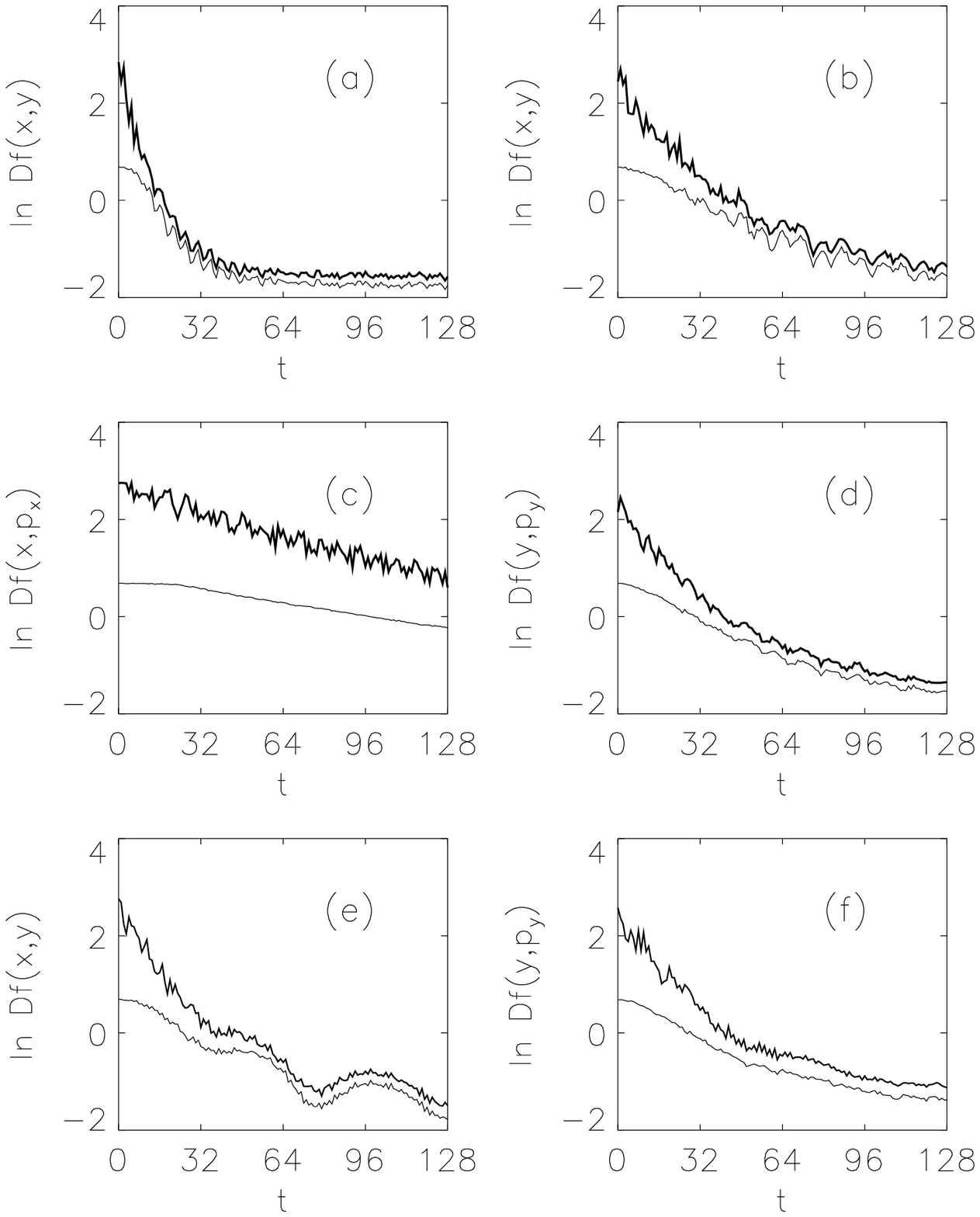}
           }
        \begin{minipage}{12cm}
        \end{minipage}
        \vskip -0.0in\hskip -0.0in
        \begin{center}\vskip .0in\hskip 0.5in
        Figure 1.
        \end{center}
\vspace{-0.2cm}
\end{figure}
\vfill\eject

\pagestyle{empty}
\begin{figure}[t]
\centering
\centerline{
        \epsfxsize=12cm
        \epsffile{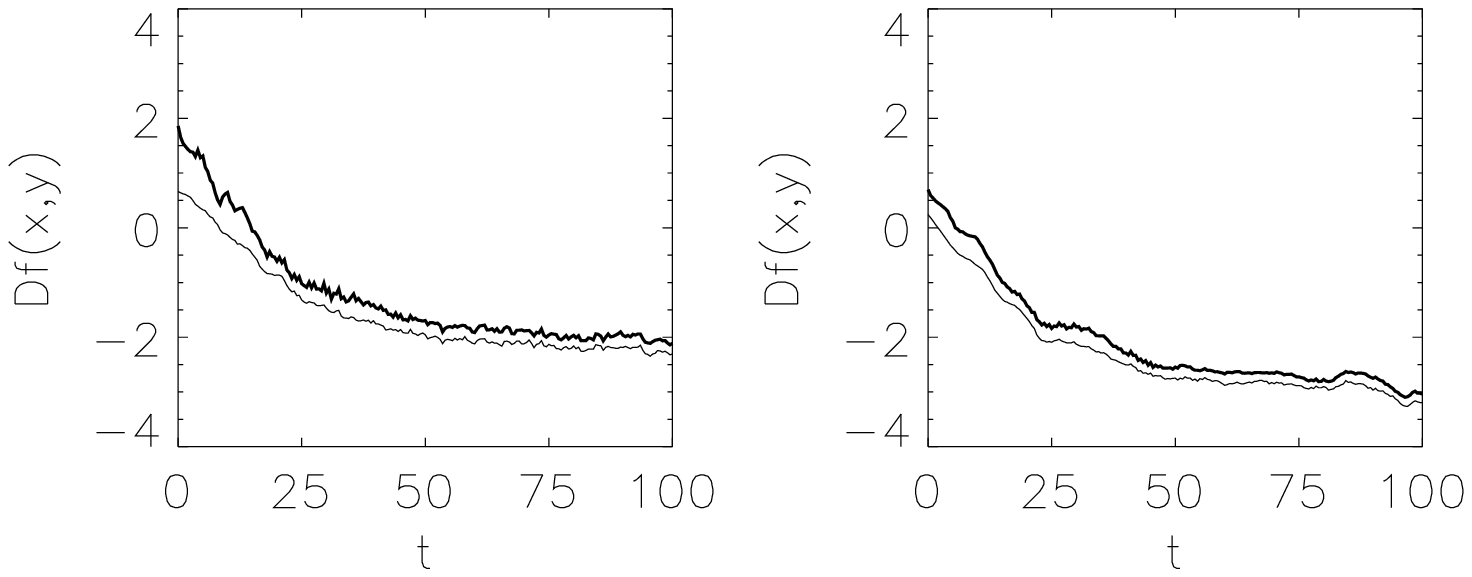}
           }
        \begin{minipage}{12cm}
        \end{minipage}
        \vskip -0.0in\hskip -0.0in
        \begin{center}\vskip .0in\hskip 0.5in
        Figure 2.
        \end{center}
\vspace{-0.2cm}
\end{figure}
\vfill\eject

\pagestyle{empty}
\begin{figure}[t]
\centering
\centerline{
        \epsfxsize=12cm
        \epsffile{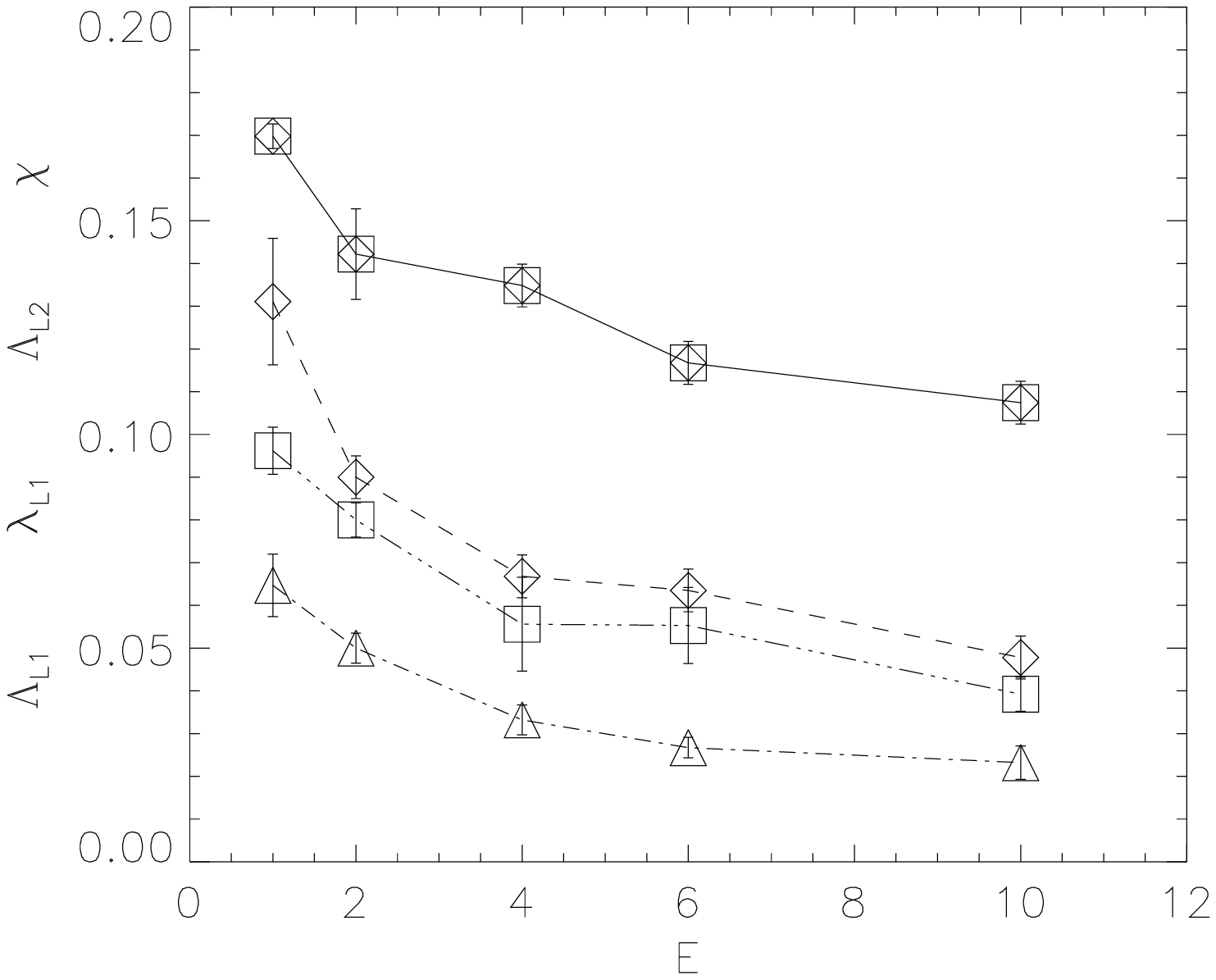}
           }
        \begin{minipage}{12cm}
        \end{minipage}
        \vskip -0.0in\hskip -0.0in
        \begin{center}\vskip .0in\hskip 0.5in
        Figure 3.
        \end{center}
\vspace{-0.2cm}
\end{figure}
\vfill\eject

\pagestyle{empty}
\begin{figure}[t]
\centering
\centerline{
        \epsfxsize=12cm
        \epsffile{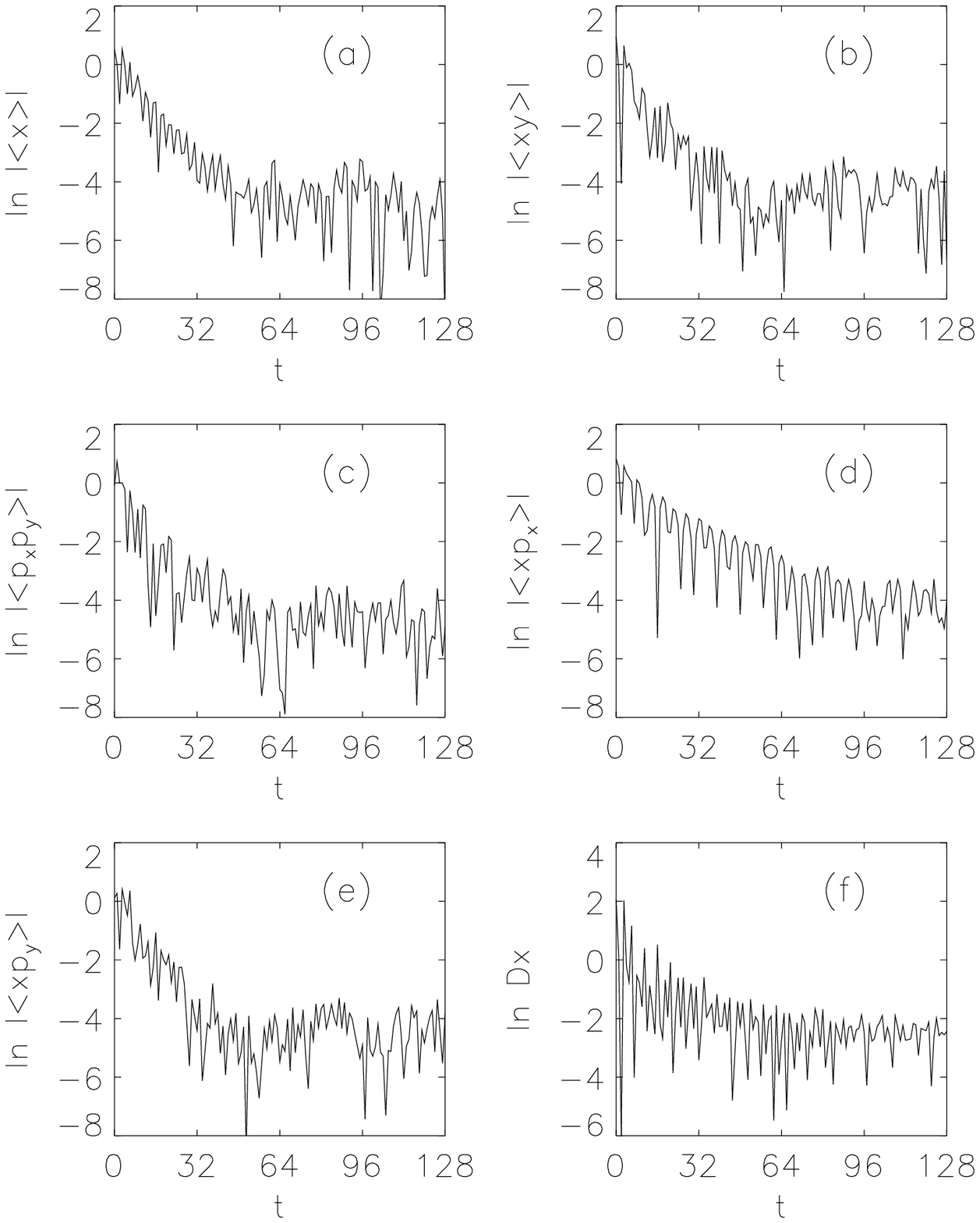}
           }
        \begin{minipage}{12cm}
        \end{minipage}
        \vskip -0.0in\hskip -0.0in
        \begin{center}\vskip .0in\hskip 0.5in
        Figure 4.
        \end{center}
\vspace{-0.2cm}
\end{figure}
\vfill\eject

\pagestyle{empty}
\begin{figure}[t]
\centering
\centerline{
        \epsfxsize=12cm
        \epsffile{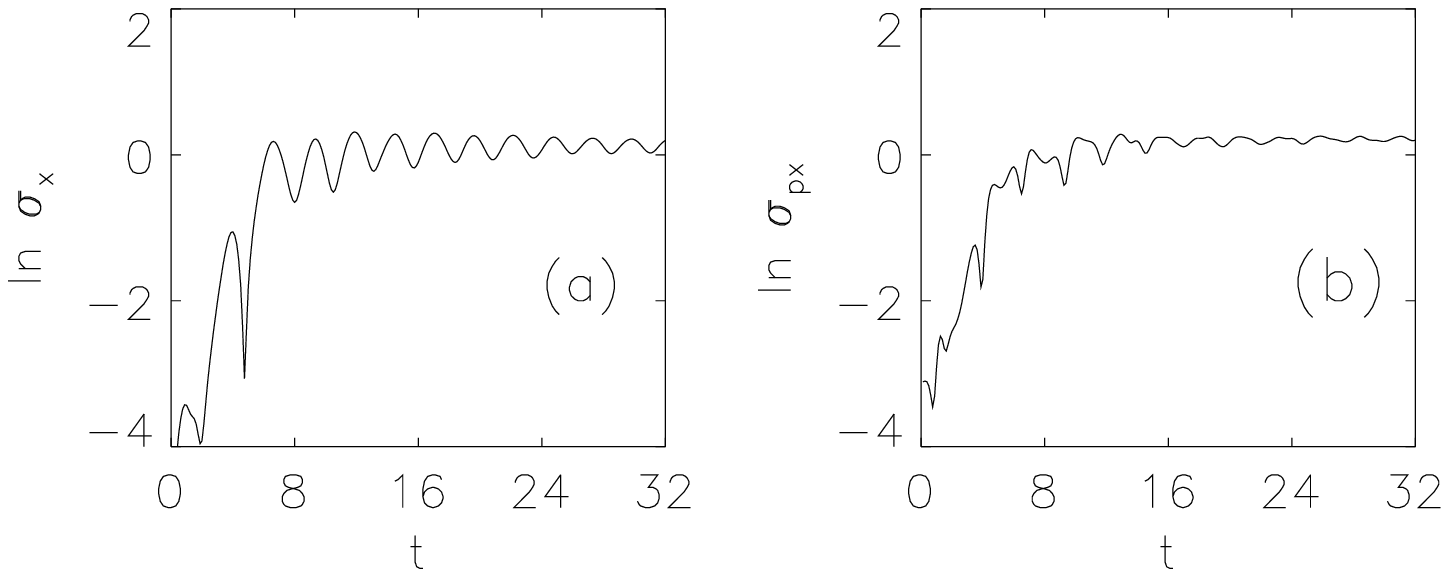}
           }
        \begin{minipage}{12cm}
        \end{minipage}
        \vskip -0.0in\hskip -0.0in
        \begin{center}\vskip .0in\hskip 0.5in
        Figure 5.
        \end{center}
\vspace{-0.2cm}
\end{figure}
\vfill\eject

\pagestyle{empty}
\begin{figure}[t]
\centering
\centerline{
        \epsfxsize=12cm
        \epsffile{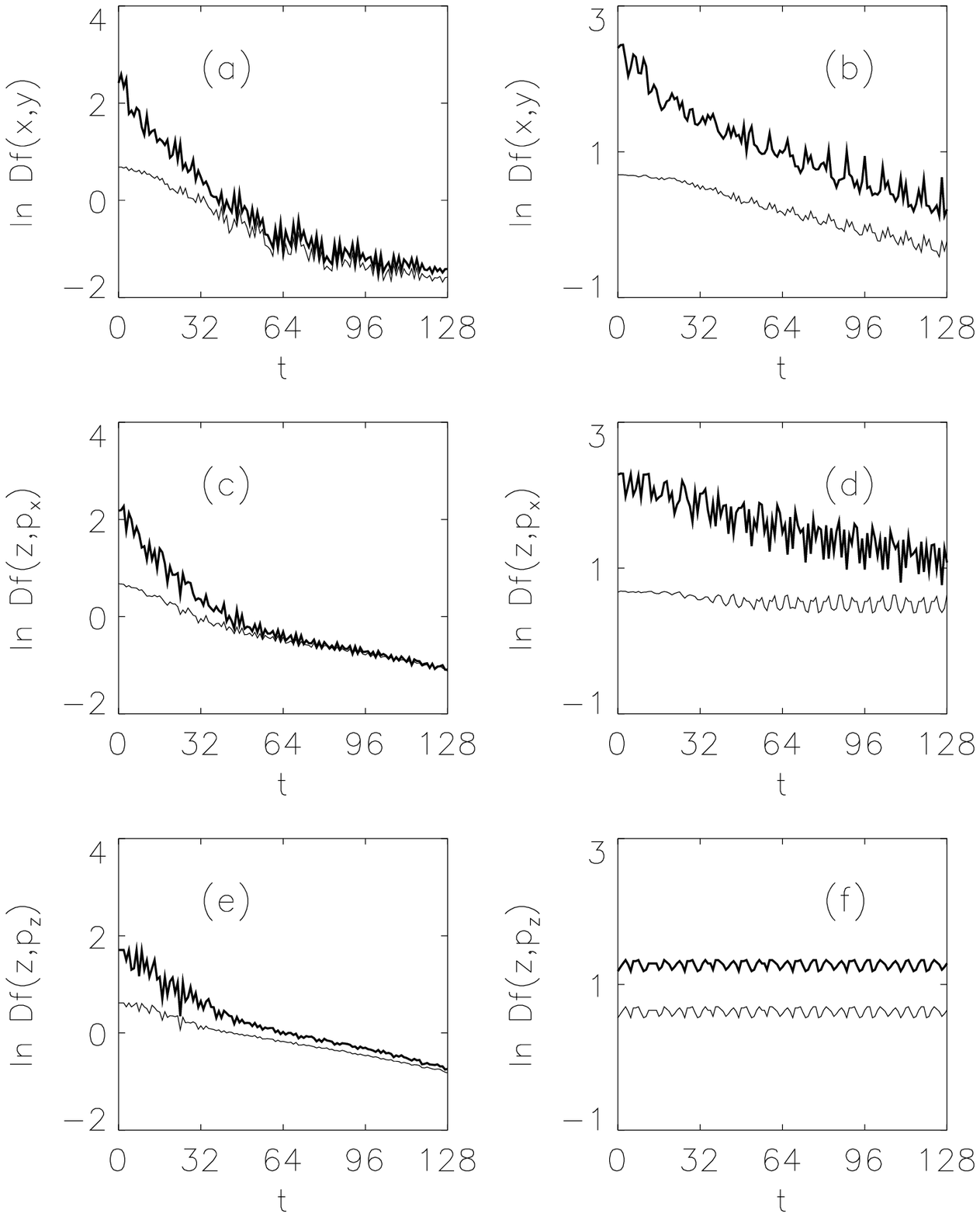}
           }
        \begin{minipage}{12cm}
        \end{minipage}
        \vskip -0.0in\hskip -0.0in
        \begin{center}\vskip .0in\hskip 0.5in
        Figure 6.
        \end{center}
\vspace{-0.2cm}
\end{figure}
\vfill\eject

\pagestyle{empty}
\begin{figure}[t]
\centering
\centerline{
        \epsfxsize=12cm
        \epsffile{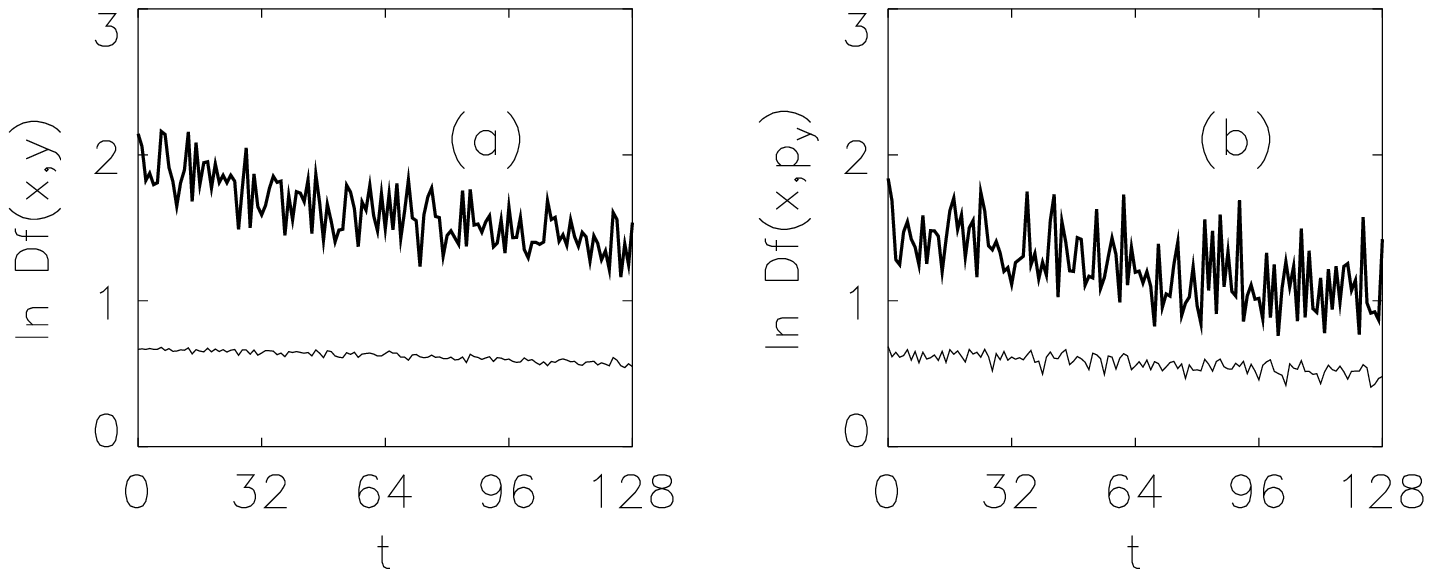}
           }
        \begin{minipage}{12cm}
        \end{minipage}
        \vskip -0.0in\hskip -0.0in
        \begin{center}\vskip .0in\hskip 0.5in
        Figure 7.
        \end{center}
\vspace{-0.2cm}
\end{figure}
\vfill\eject

\pagestyle{empty}
\begin{figure}[t]
\centering
\centerline{
        \epsfxsize=12cm
        \epsffile{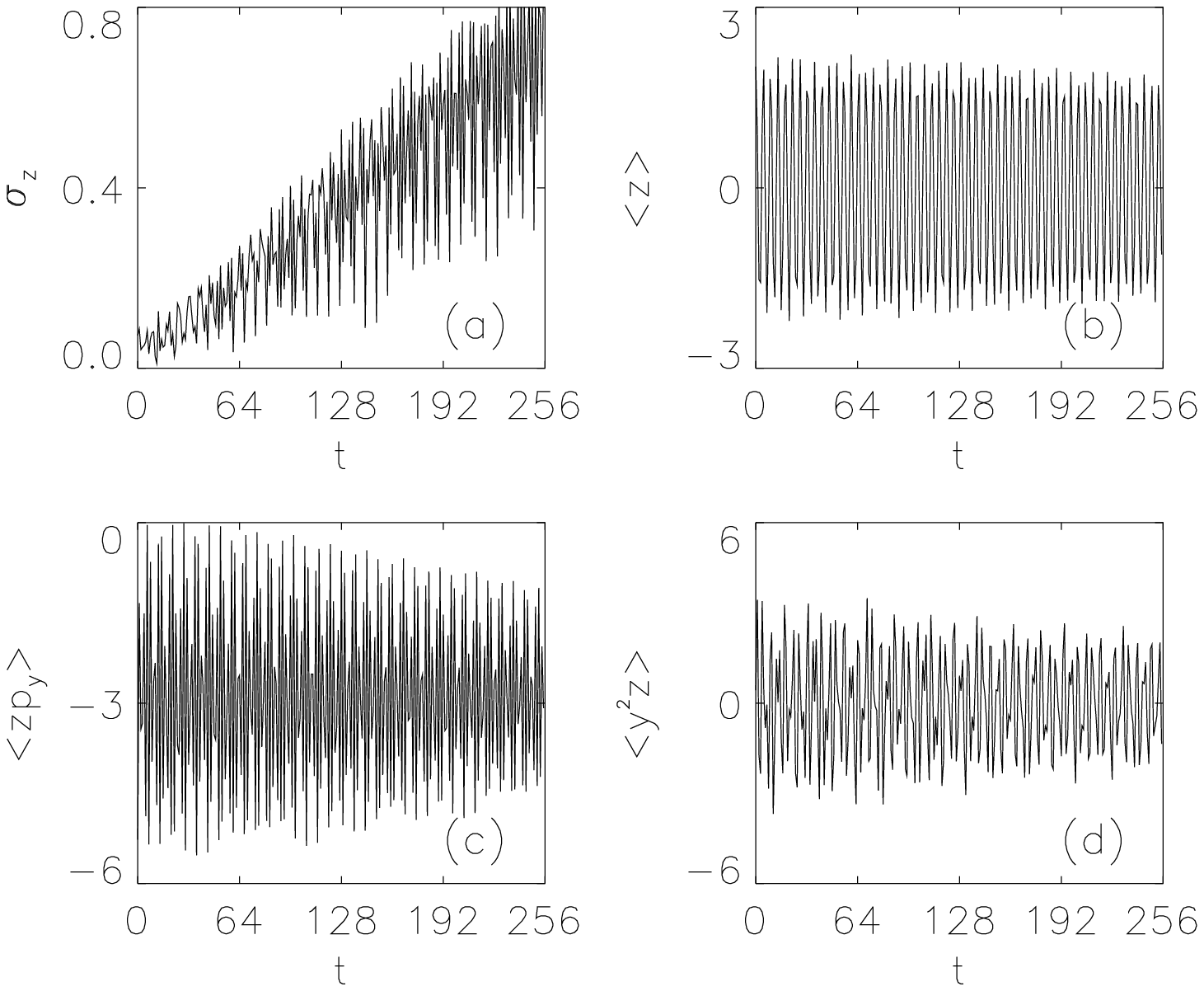}
           }
        \begin{minipage}{12cm}
        \end{minipage}
        \vskip -0.0in\hskip -0.0in
        \begin{center}\vskip .0in\hskip 0.5in
        Figure 8.
        \end{center}
\vspace{-0.2cm}
\end{figure}
\vfill\eject

\pagestyle{empty}
\begin{figure}[t]
\centering
\centerline{
        \epsfxsize=12cm
        \epsffile{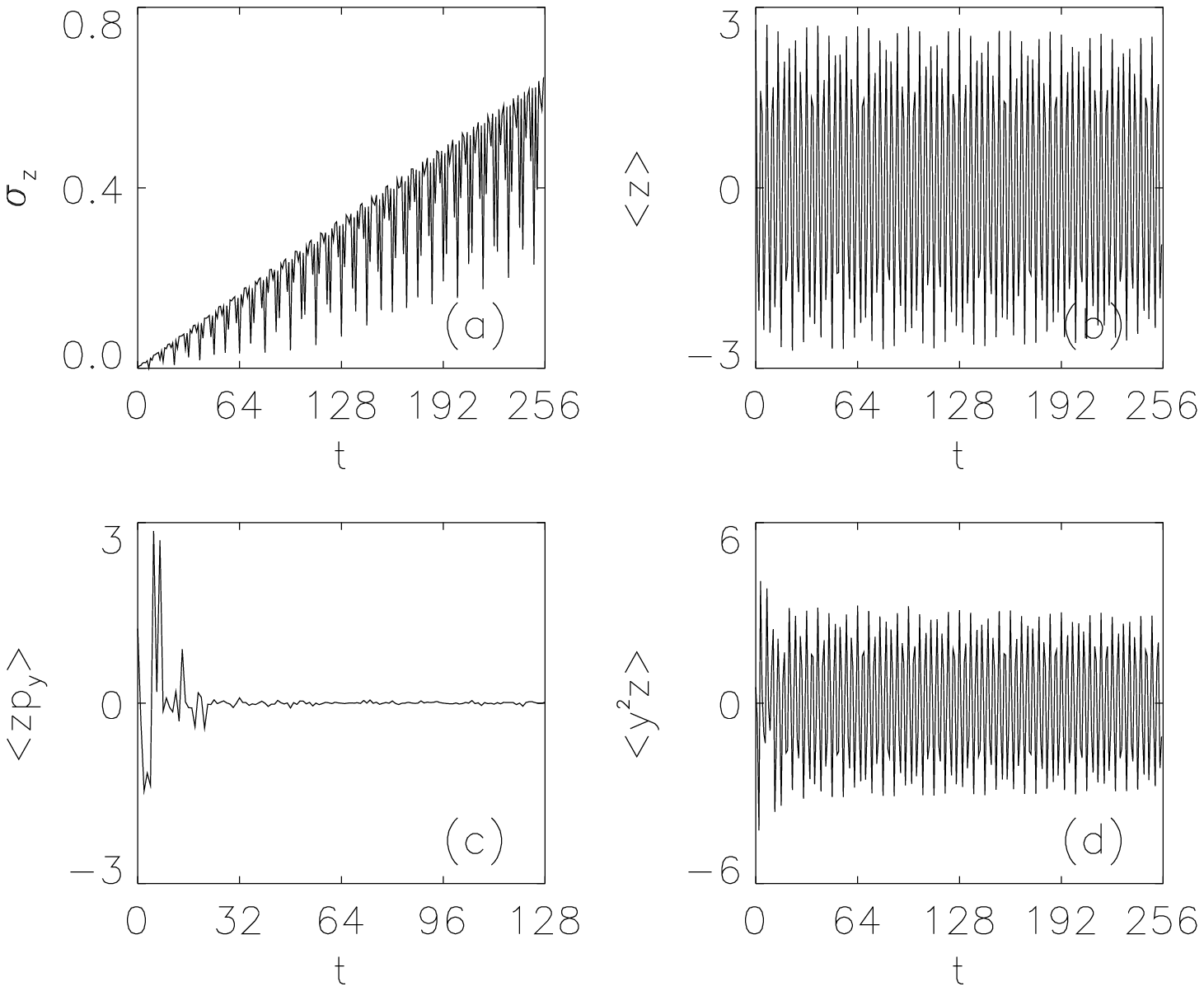}
           }
        \begin{minipage}{12cm}
        \end{minipage}
        \vskip -0.0in\hskip -0.0in
        \begin{center}\vskip .0in\hskip 0.5in
        Figure 9.
        \end{center}
\vspace{-0.2cm}
\end{figure}
\vfill\eject

\pagestyle{empty}
\begin{figure}[t]
\centering
\centerline{
        \epsfxsize=12cm
        \epsffile{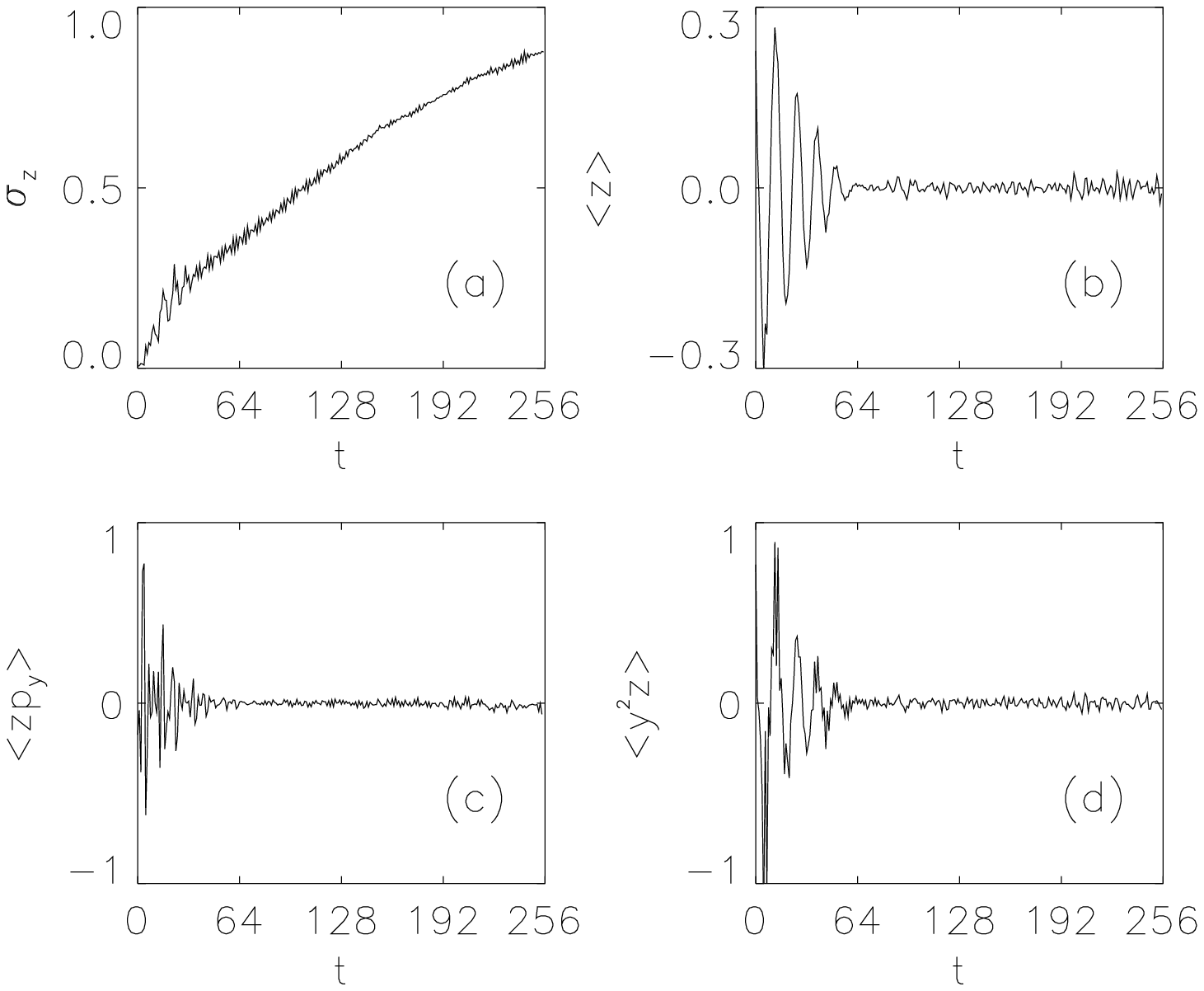}
           }
        \begin{minipage}{12cm}
        \end{minipage}
        \vskip -0.0in\hskip -0.0in
        \begin{center}\vskip .0in\hskip 0.5in
        Figure 10.
        \end{center}
\vspace{-0.2cm}
\end{figure}
\vfill\eject

\pagestyle{empty}
\begin{figure}[t]
\centering
\centerline{
        \epsfxsize=12cm
        \epsffile{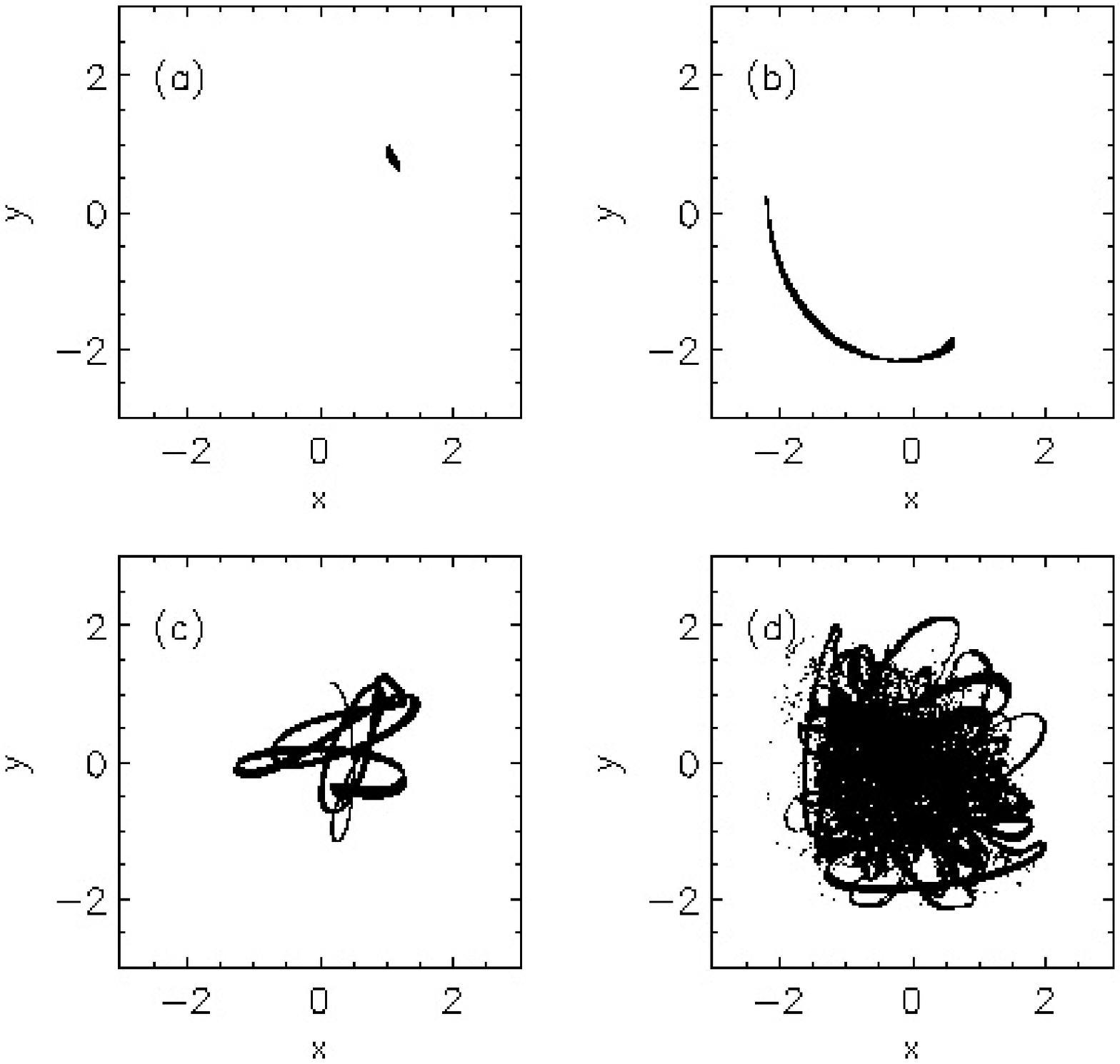}
           }
        \begin{minipage}{12cm}
        \end{minipage}
        \vskip -0.0in\hskip -0.0in
        \begin{center}\vskip .0in\hskip 0.5in
        Figure 11.
        \end{center}
\vspace{-0.2cm}
\end{figure}
\vfill\eject
\end{document}